\documentclass[preprint]{revtex4}

\usepackage{amsmath, amssymb}
\usepackage{natbib}
\usepackage[colorlinks, 
linkcolor=blue,
urlcolor=blue,
citecolor=blue,
bookmarksopen=false]{hyperref}

\usepackage{amsmath}
\usepackage{graphicx, xcolor}
\usepackage{caption}
\usepackage{subcaption}

\newcommand{\fsa}[1]{\langle{#1}\rangle}
\def\iotabar{\lower3pt\hbox{$\mathchar'26$}\mkern-7mu\iota}
\newcommand {\aplt} {\ {\raise-.5ex\hbox{$\buildrel<\over\sim$}}\ }
\newcommand{\dd}{\mbox{d}}

\newcommand{\eq}[1]{(\ref{#1})}
\newcommand{\bun}{\hat{\mathbf{b}}}

\newcommand{\bB}{\mathbf{B}}

\newcommand{\bx}{\mathbf{x}}

\newcommand{\dotcross}{ \raise 0.65ex\hbox{${\scriptstyle {{_{\displaystyle \cdot}}\atop\times}}$} }
\newcommand{\crossdot}{ \raise 0.5ex\hbox{${\scriptstyle {{_\times}\atop{\displaystyle \cdot}}}$} }

\newcommand{\sumsig}{ \raise -1.3ex\hbox{${{\displaystyle \sum}\atop{\scriptstyle \sigma}}$} }
\newcounter{appnumb}

\begin{document}

\title{Robust stellarator optimization via flat mirror magnetic fields}
\author{J.~L. Velasco$^1$, I. Calvo$^1$, E. S\'anchez$^1$ and F.I. Parra$^2$}
\affiliation{$^1$Laboratorio Nacional de Fusi\'on, CIEMAT, Madrid, {28040}, Spain\\ $^2$Princeton Plasma Physics Laboratory, Princeton, NJ 08543, USA}

\date{\today}

\begin{abstract}
Stellarator magnetic configurations need to be optimized in order to meet all the required properties of a fusion reactor. In this work, it is shown that a flat-mirror quasi-isodynamic configuration (i.e. a quasi-isodynamic configuration with sufficiently small radial variation of the mirror term) can achieve small radial transport of energy and good confinement of bulk and fast ions even if it is not very close to perfect omnigeneity, and for a wide range of plasma scenarios, including low $\beta$ and small radial electric field. This opens the door to constructing better stellarator reactors. On the one hand, they would be easier to design, as they would be robust against error fields. On the other hand, they would be easier to operate since, both during startup and steady-state operation, they would require less auxiliary power, and the damage to plasma-facing components caused by fast ion losses would be reduced to acceptable levels.
\end{abstract}

\maketitle

\section{Introduction}\label{SEC_INTRO}

Stellarators are an alternative to tokamaks in the quest for a fusion reactor based on magnetic confinement. Their most prominent advantage is that their magnetic field is created almost entirely by external coils. No time-dependent current is thus needed inside the plasma, which enables steady-state operation and makes stellarators less prone to large-scale instabilities than tokamaks. On the other hand, the magnetic configuration of the stellarator is three-dimensional, which generally worsens their neoclassical confinement (the one associated to the inhomogeneity of the magnetic field and inter-particle collisions) of both reactants and fusion-born alpha particles to a level incompatible with reactor operation. This has historically left the stellarator concept at a disadvantage with respect to the tokamak, whose axial symmetry makes neoclassical transport negligible at the low collisionalities relevant for the core of a reactor.

The strategy employed to mitigate this problem is stellarator optimization: the magnetic configuration is carefully tailored in order to have (among other criteria) a neoclassical transport as low as that of the axisymmetric tokamak. A stellarator configuration that fulfils this criterion is called~\textit{omnigeneous}~\cite{cary1997omni}. Quasi-isodynamic (QI) and quasi-symmetric configurations (QS) are specific types of omnigenous stellarators, see e.g.~\cite{landreman2012omni}.

\begin{figure}
\begin{center}
\includegraphics[angle=0,width=\columnwidth]{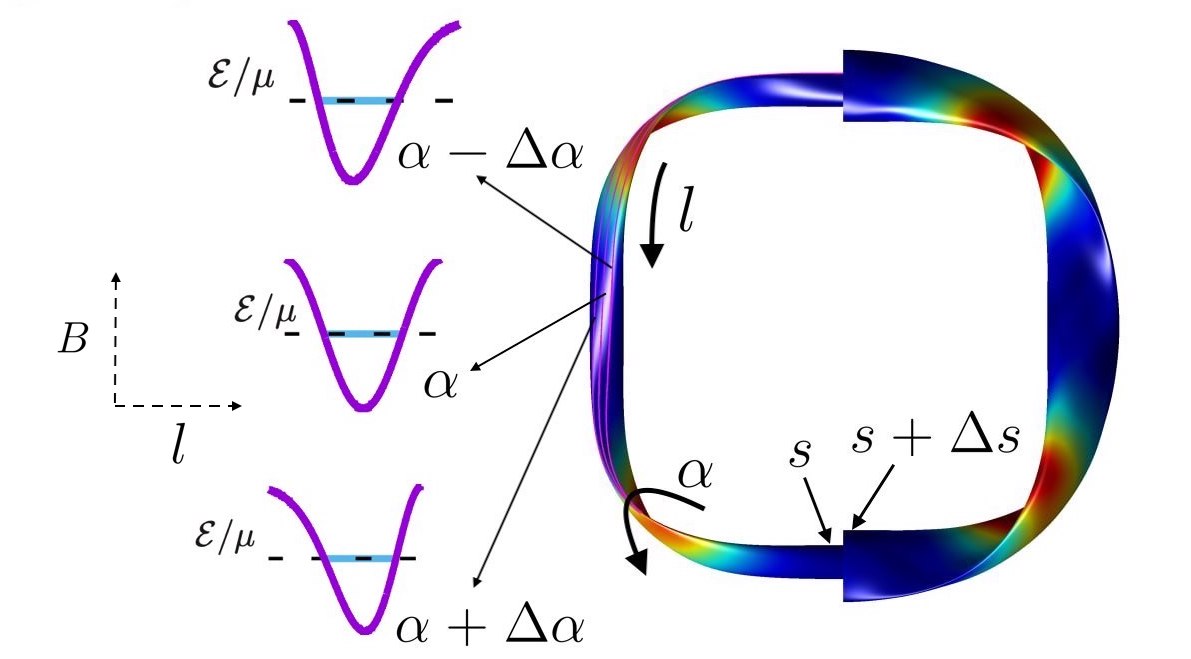}  
\end{center}
\caption{Top view of two flux-surfaces of a stellarator magnetic configuration optimized to be quasi-isodynamic. The color indicates magnetic field strength (red is large, blue is small). Part of three magnetic field lines on the same flux-surface is highlighted in magenta. The sketches illustrate the variation of the magnetic field strength along these field lines.}
\label{FIG_CIEMATQI}
\end{figure}

QI magnetic fields are omnigeneous fields in which the contours of constant magnetic field strength, $B = |\bB|$, close in the poloidal direction. An example is shown in figure \ref{FIG_CIEMATQI}, which represents $B$ on two flux-surfaces of a nearly QI stellarator configuration~\cite{sanchez2023qi}. QI devices are sometimes called \textit{linked mirrors}~\cite{wobig1993helias}. When viewed from above, configurations that approach quasi-isodynamicity have a polygonal shape (a square, in this particular case), with the maxima of $B$ (red) being usually largest in the corner regions. The particles with largest contribution to the energy losses reside in the relatively straight sections of small $B$ (blue), moving back and forth along the magnetic field line (part of three field lines is highlighted in magenta in figure \ref{FIG_CIEMATQI}), hence the term \textit{mirror}. On the left of figure \ref{FIG_CIEMATQI}, three sketches represent $B$ along these field lines on the same flux-surface in one of the straight sections. On a longer time scale, these \textit{trapped} particles \textit{drift} in the direction perpendicular to the magnetic field. In an axisymmetric tokamak, on average, the drift velocity is tangent to the magnetic surface, which is at the root of all its good properties with respect to neoclassical transport. However, in a generic stellarator configuration, the drift velocity has a component perpendicular to the flux surface, which leads to large particle and energy losses. In an exactly QI stellarator, the component of the drift velocity perpendicular to the flux surface vanishes, as in the tokamak, thanks to a careful design of the nearby \textit{mirrors} of the same flux-surface. When drifting on the flux-surface (which means going from top to bottom in the sketches of figure \ref{FIG_CIEMATQI}), all trapped particles in a QI field need to encounter a mirror field whose maximum and minimum values do not vary, and such that the distance between the tips of its trajectory does not change (the specific properties of such a field have been derived in \cite{cary1997omni}).

The experiment Wendelstein 7-X (W7-X) is the main representative of the success of this approach. It has demonstrated that a large optimized stellarator with superconducting coils can be built~\cite{sunnpedersen2016nature} and operated~\cite{klinger2017op11,klinger2019op12,sunnpedersen2022w7x}, leading to record performance~\cite{klinger2019op12} thanks precisely to its optimization with respect to neoclassical transport~\cite{beidler2021nature}. Nevertheless, its neoclassical confinement of fast ions is expected to be good only at high plasma pressure (while reactor operation makes this property desirable already at low plasma pressure), and only for ions born sufficiently close to the magnetic axis~\cite{drevlak2014fastions}. Moreover, it is worse than in earlier design configurations due to modifications of the coil geometry~\cite{drevlak2014fastions}. Additionally, the level of neoclassical transport of energy is still relatively large in the high-mirror magnetic configuration of W7-X, the one that, if extrapolated to a reactor, would be compatible with the operation of an island divertor~\cite{sunnpedersen2019divertor}. Furthermore, an additional problem concerns the energy transport caused by turbulent fluctuations in the plasma. Traditionally overlooked in stellarator optimization, due to its computational cost and to the importance of neoclassical transport, it is now recognized that turbulent transport must be minimized in order to achieve high temperature stellarator plasmas~\cite{carralero2021dr,carralero2022dr}. On the other hand, some W7-X plasmas with reduced turbulence present neoclassical impurity transport that causes accumulation~\cite{romba2023nbi}, a problem that, as we will discuss later, may be exacerbated by its optimization strategy with respect to bulk energy transport. Finally, it should be noted that fabricating and assembling the three-dimensional non-planar coils of W7-X was a challenge~\cite{bosch2013challenges}. 

It is therefore clear that a more efficient and robust optimization of the transport of bulk and energetic species, if possible, could be very helpful for the design of the next generation of stellarators. With this purpose, in this work we introduce the concept of flat mirror quasi-isodynamic stellarators. We will show that robustly optimized stellarator configurations can be obtained if, in addition to the tailoring of the mirrors \textit{within} the flux-surfaces of the device, special care is taken of the variation of the shape of these mirrors \textit{across} the different flux-surfaces. Specifically, if the radial variation of the amplitude of the mirror is small, hence the term 'flat mirror'.

The rest of the paper is organized as follows. In section \ref{SEC_PICTURE}, we give a physical picture for the need for robust optimization and for the general strategy to achieve it. In section~\ref{SEC_THEORY}, we provide the required theoretical background: we discuss particle orbit optimization in terms of the second adiabatic invariant $J$, and show that a negative radial derivative of $J$ has beneficial effects for transport. In section~\ref{SEC_DSJ}, we present a strategy for designing stellarators with this property. We assess this strategy in sections~\ref{SEC_FASTIONS} and \ref{SEC_BULKIONS}, for fast and bulk ions respectively, by analyzing the optimization campaign of the magnetic configuration CIEMAT-QI~\cite{sanchez2023qi}. The conclusions come in section~\ref{SEC_DISCUSSION}. Appendices~\ref{SEC_CALC},~\ref{SEC_B0} and~\ref{SEC_BOTTOM} contain the derivation and justification of the analytical results employed along this paper.

\section{Physical picture of robust optimization}\label{SEC_PICTURE}

It is important to explain the limitations of addressing the design of a stellarator configuration by exclusively aiming for closeness to omnigeneity, without including (for example) the robustness criteria that we discuss in this paper. We will do that in this section, in which we will draw a first connection between the robustness of an optimization and the radial variation of the \textit{system of linked mirrors}. This connection will be rigorously made in sections \ref{SEC_THEORY} and \ref{SEC_DSJ}.

The spatial coordinates employed in this work are depicted in Figure \ref{FIG_CIEMATQI}. We use $s$, a flux-surface label (for the calculations we will employ $s=\Psi/\Psi_{LCFS}$, where $2\pi\Psi$ is the toroidal flux through the flux-surface, and $\Psi=\Psi_{LCMS}$ at the last closed flux-surface), $\alpha$, a field line label (in this work, $\alpha=\theta-\iota\zeta$, with $\theta$ and $\zeta$ poloidal and toroidal Boozer angles, and $\iota$ the rotational transform), and $l$, the arc-length along the magnetic field line. The magnetic field can be written as
\begin{equation} 
\mathbf{B}=\partial_s\Psi(s)\nabla s\times\nabla\alpha\,.
\end{equation}
Trapped orbits are orbits of particles whose parallel velocity $v_\parallel = \mathbf{v}\cdot \bB/B$ becomes zero at some point of their trajectory and thus bounce back and forth along $l$ while slowly drifting across the magnetic field. In the absence of collisions, the energy
\begin{equation} 
\mathcal{E}=\frac{1}{2}v^2+\frac{Ze\Phi}{m}
\end{equation}
and the magnetic moment 
\begin{equation} 
\mu=\frac{v^2-v_\parallel^2}{2ZeB}
\end{equation}
(both defined per mass unit) are conserved. Here, $v^2=\mathbf{v}\cdot\mathbf{v}$, $\Phi$ is the electrostatic potential, and $m$ and $Ze$ are the particle mass and charge respectively. Trapped orbits can be characterized by the second adiabatic invariant
{\begin{equation} 
J(s,\alpha,\mathcal{E},\mu)\equiv 2\int_{l_{b_1}}^{l_{b_2}}\mathrm{d}l |v_\parallel |= 2\int_{l_{b_1}}^{l_{b_2}}\mathrm{d}l \sqrt{2\left(\mathcal{E}-\mu B-\frac{Ze\Phi}{m}\right)}\,.
\end{equation} 
Here, the integral over $l$ is taken between the bounce points $l_{b_1}$ and $l_{b_2}$, i.e., between the points where the parallel velocity becomes zero. In the absence of collisions, particles move at constant $J$.

\begin{figure}
\begin{center}
\includegraphics[angle=0,width=0.49\columnwidth]{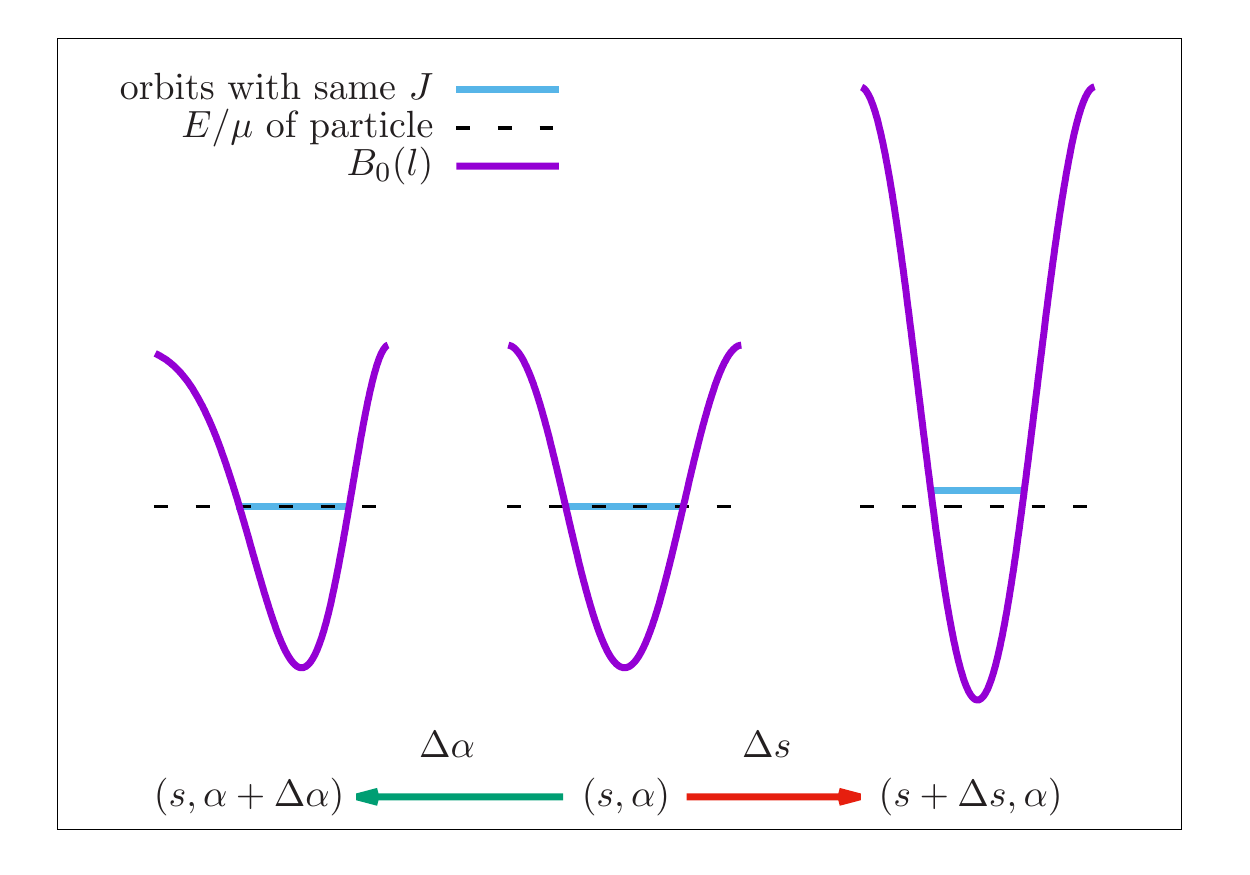}
\includegraphics[angle=0,width=0.49\columnwidth]{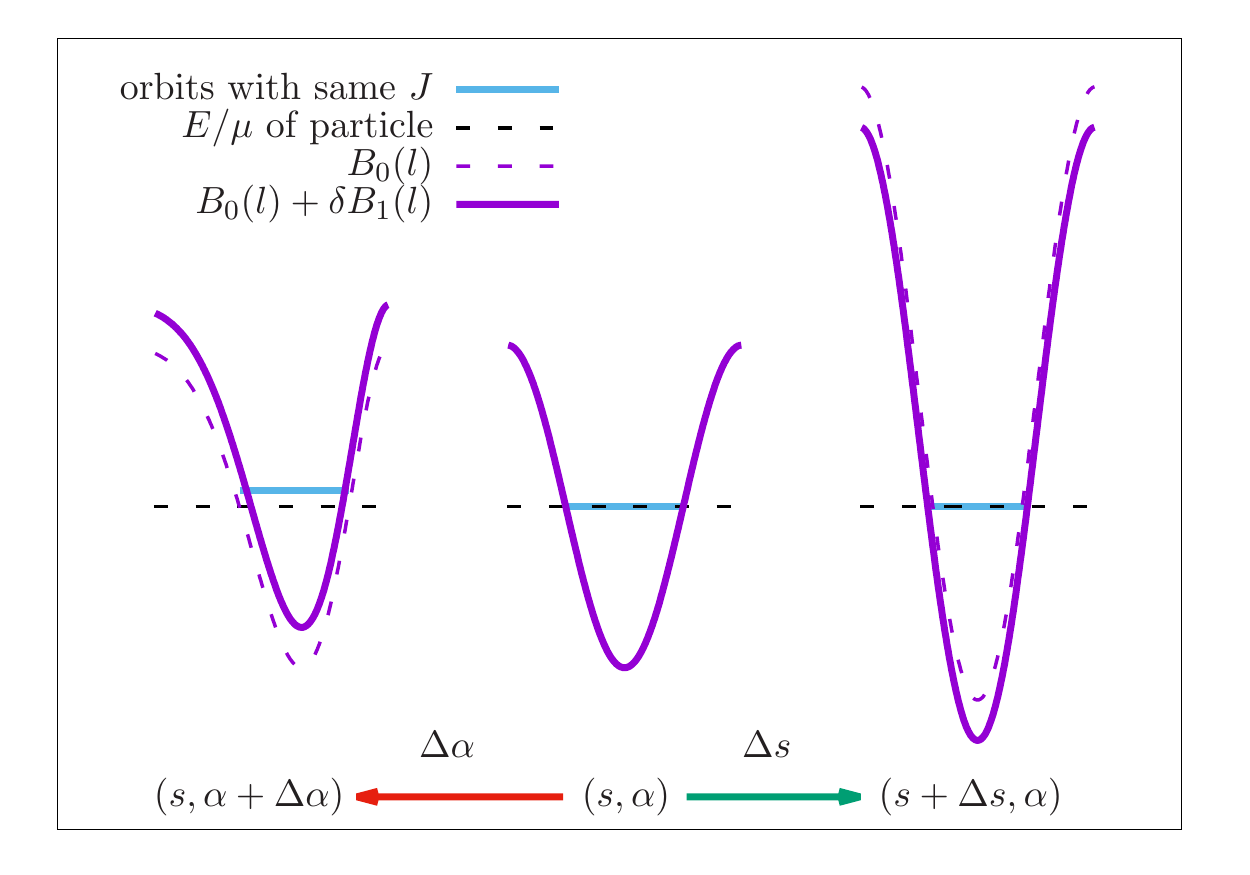}
\includegraphics[angle=0,width=0.49\columnwidth]{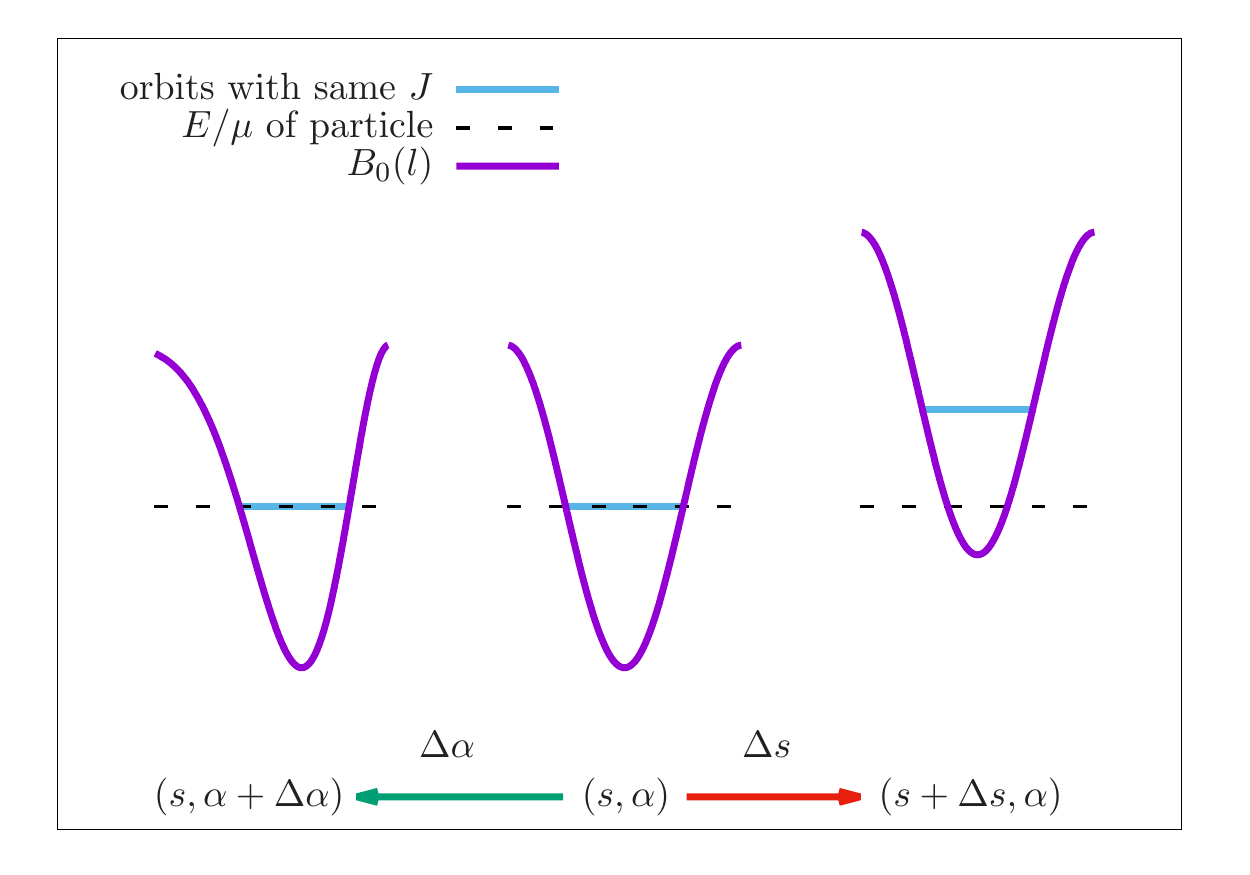}
\includegraphics[angle=0,width=0.49\columnwidth]{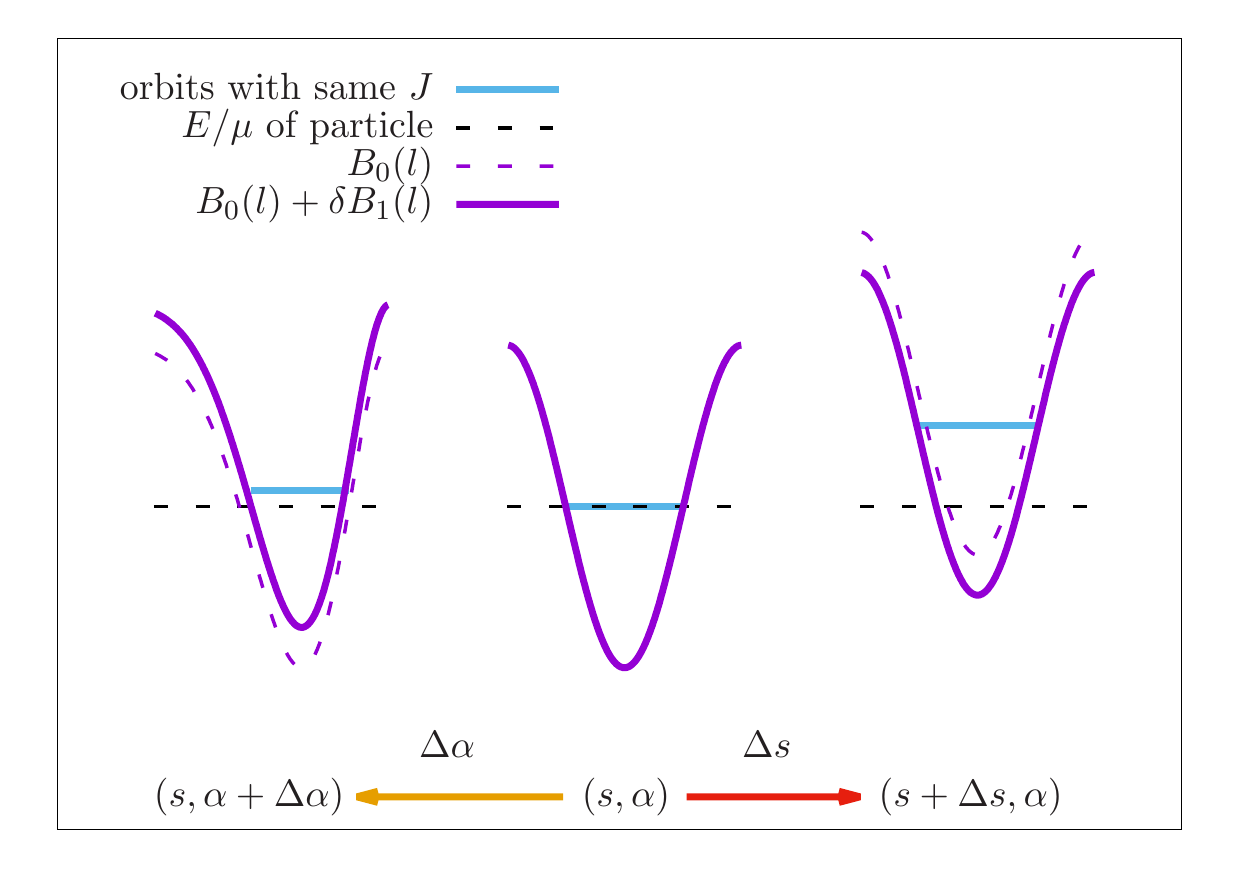}
\end{center}
\caption{Sketch of mirror fields and particle trajectories in an exactly QI field (top left), a perturbed QI field (top right), an exactly QI field with a flat mirror (bottom left) and a perturbed QI field with a flat mirror (bottom right). Purple lines indicate the magnetic field strength along the field line (dashed purple lines represent the original QI field in the right column). The dashed black lines mark constant $\cal{E}/\mu$ and orbits with the same $J$ are highlighted in light blue. For each configuration, trajectories are depicted at ($s,\alpha$) (center of the subfigure), ($s, \alpha+\Delta\alpha$) (left) and ($s+\Delta s, \alpha$) (right).}\label{FIG_MIRRORS}
\end{figure}


Let us start by assuming that the magnetic field strength $B$ of a configuration sufficiently close to QI can be split into a QI piece $B_0$ and a non-QI piece $B_1$, with
\begin{equation}
{B}={B_0}+\delta{B_1}\,,\quad {B_0}\sim{B_1}\,,\quad 0\leq\delta\ll 1.
\label{EQ_SPLITTING}
\end{equation}
Here, $\delta=0$ (and thus $B=B_0$) for an exactly QI field. Figure \ref{FIG_MIRRORS} contains sketches of mirror fields $B$ and particle trajectories at several locations of different magnetic configurations. They are to be interpreted as follows: a trapped particle at ($s,\alpha$) (center of each subfigure) with energy and magnetic moment per mass unit $\mathcal{E}$ and $\mu$, is moving back and forth along $l$ (light blue) in an orbit characterized by $J$. On a longer time scale, it will move at constant $\mathcal{E}/\mu$ (i.e., horizontally in the figure) and constant $J$ (i.e., to another light blue segment) in the radial direction ($\Delta s$), in the direction tangent to the flux-surface ($\Delta \alpha$), or in a combination of both. By definition, in an exactly QI field (that we denote $B_0$) $J$ does not depend on the field line if everything else is left constant~\cite{cary1997omni}. This is indicated in figure \ref{FIG_MIRRORS} (left column) by the fact that the light blue orbits at ($s, \alpha$) and ($s, \alpha+\Delta\alpha$) correspond to the same value of $\mathcal{E}/\mu$. This is not the case in magnetic fields that are not exactly QI, as in figure \ref{FIG_MIRRORS} (right column).

In a QI field such as the one of figure \ref{FIG_MIRRORS} (top left), a particle moving at constant $\mathcal{E}/\mu$ will drift tangentially towards ($s, \alpha+\Delta\alpha$). Meanwhile, the motion towards ($s+\Delta s, \alpha$) at constant $\mathcal{E}/\mu$ is forbidden, as it would not conserve $J$. Things may change if, on top of the exactly QI field, a perturbation $\delta B_1$ is added. In the case of figure \ref{FIG_MIRRORS} (top right), the perturbation is such that the particle, moving at constant $\mathcal{E}/\mu$ and $J$, will now drift in the radial direction. It can be inferred from figures \ref{FIG_MIRRORS} (top row) that this is the consequence of $J$ having a small radial variation: as soon as a perturbation is added to the exactly QI field, $J$ at $s+\Delta s$ becomes comparable to $J$ at $s$. The role of the radial variation of $J$ on several aspects of stellarator transport will be the discussed in detail in section \ref{SEC_THEORY}.

This lack of robustness can be solved, as we show in figure \ref{FIG_MIRRORS} (bottom), by designing the QI field at $s+\Delta s$ to be very different from the QI field at $s$, specifically via a mirror term whose shape and size does not change radially. The scenario in figure \ref{FIG_MIRRORS} (bottom left) is nearly identical to that of figure \ref{FIG_MIRRORS} (top left): for an exactly QI field, no matter its radial variation, the drift of the trapped particles will take place in the poloidal direction. But now this situation is not qualitatively modified when the perturbation is added in figure \ref{FIG_MIRRORS} (bottom right): radial drift is still forbidden by conservation of $J$, and the particle will drift in a direction that is almost tangent to the flux-surface, because $J$ at ($s, \alpha+\Delta\alpha$) is still relatively close to $J$ at ($s, \alpha$). This is enough to guarantee, for instance, good fast ion confinement \cite{velasco2021prompt}. The key for this result has been that the configuration has what in this work we call a (radially) flat mirror: the mirror term has a small radial variation and, as a consequence of this, the magnetic field at $s+\Delta s$ in figure \ref{FIG_MIRRORS} (bottom left) has roughly the same amplitude that it has at $s$, but it is displaced by an offset. While exactly the same offset was present in figure \ref{FIG_MIRRORS} (top left), the mirror amplitude had a stronger radial variation in that case, resulting in low robustness of the stellarator optimization. The specific characteristics of a such flat-mirror QI field will be presented in section \ref{SEC_DSJ}, in equation (\ref{EQ_FLATMIRROR}).

\begin{figure}
\begin{center}
\includegraphics[angle=90,width=0.49\columnwidth]{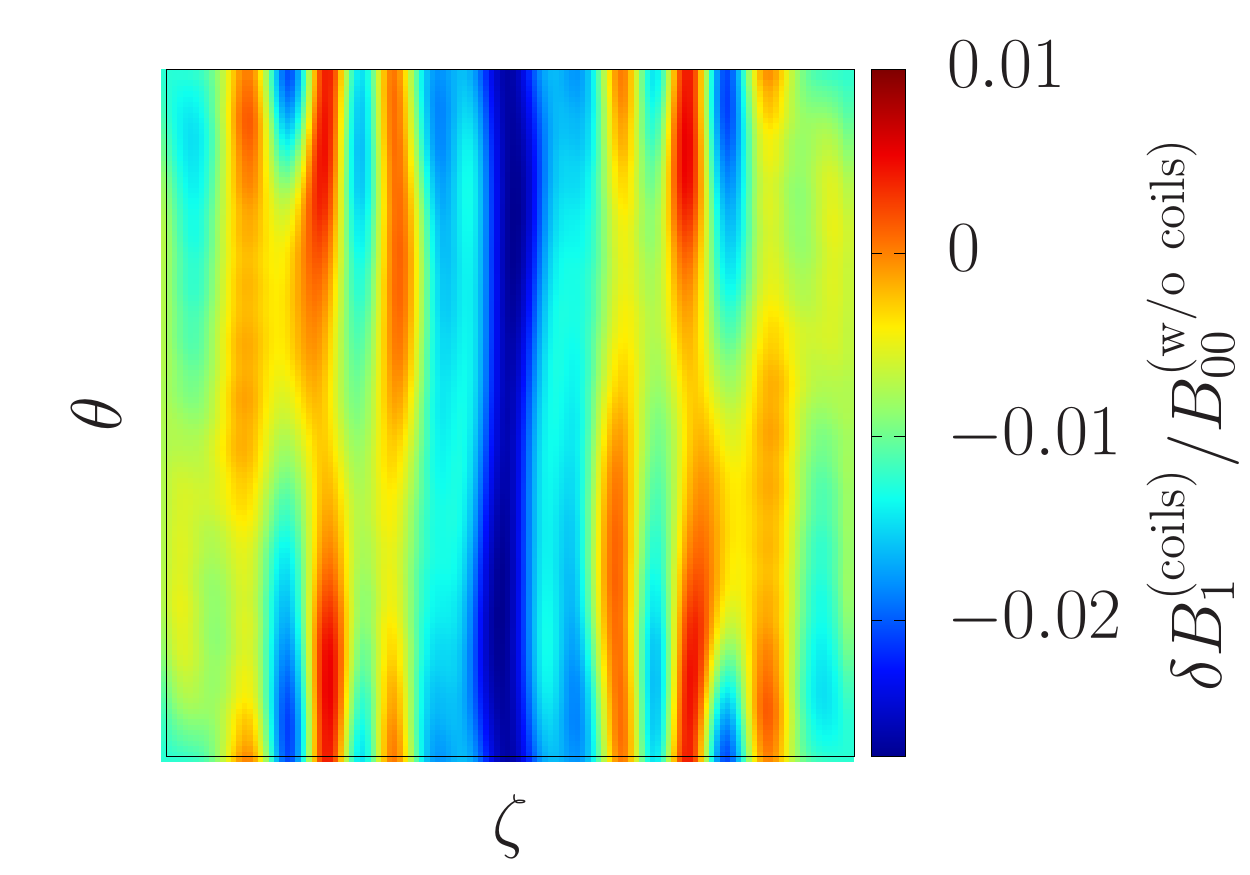}
\includegraphics[angle=90,width=0.49\columnwidth]{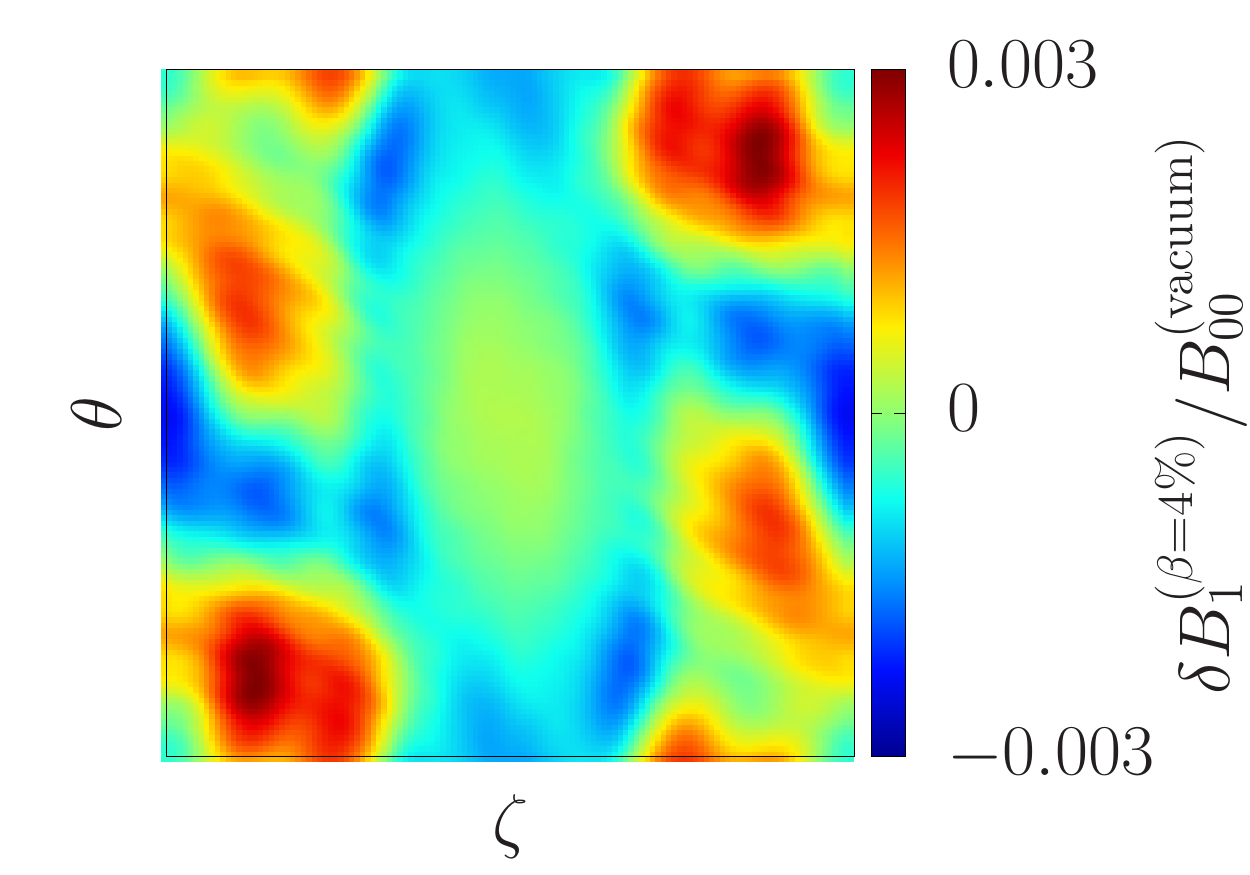}
\end{center}
\caption{Difference between the magnetic field strength of CIEMAT-QI in one field-period of flux-surface $s=0.1$, generated with and without coils (left); in the same domain, change in the magnetic field strength between $\fsa{\beta}=4\%$ and $\fsa{\beta}=0\%$ (right). $\delta B_1^{\fsa{\beta}=4\%}$ is defined $\delta B_1^{\fsa{\beta}=4\%}= (B-B_{00})^{\fsa{\beta}=4\%}- (B-B_{00})^\mathrm{(vacuum)}$, where $B_{00}$ is the constant term in the Fourier expansion of $B$.\label{FIG_DELTA}}
\end{figure}

We end this section by illustrating that deviations $\delta B_1$ from an exactly omnigeneus $B_0$ are unavoidable and arise from a variety of sources. In practice, imposing the constraints given by the MHD equilibrium equations always leads to stellarator magnetic configurations that are not exactly omnigeneous. This departure from omnigeneity may be exacerbated by conflicts with other design criteria. Although an unambiguous splitting of a given $B$ into $B_0$ and $\delta B_1$ is not straightforward, order-of-magnitude estimates of $\delta$ can be drawn by inspection of the values of $B$ on the flux-surface or of its Fourier decomposition in Boozer coordinates. They range from $\delta\sim 10^{-1}$, for the high-mirror configuration of W7-X, to record values of $\delta\sim 10^{-4}$, corresponding to recently obtained precise QS fields~\cite{landreman2022preciseQS}.

Additionally, in the case of the so-called two-stage stellarator optimization strategy (see e.g. \cite{henneberg2021coils} and references therein), the second stage consists of finding a set of coils that are able to reproduce the magnetic field obtained on a first stage. This was, in particular, the strategy for CIEMAT-QI \cite{sanchez2023qi}, and figure \ref{FIG_DELTA} (left) represents the deviation between the magnetic field strength at $s=0.1$ resulting from the first stage and one with similar properties that can be obtained by means of filamentary coils~\cite{sanchez2023qi}. A deviation from omnigeneity $\delta B_1$ with $\delta\sim 10^{-2}$ can be roughly attributed to these particular coils (the reason why this is a rough estimate is that the difference between $B$ with and without coils might contain an omnigeneous piece, that strictly speaking should not be included in $\delta B_1$). On top of this, displacements or deformations of the final coils, once the device is built, may cause error fields, see \cite{lazerson2018error} and references therein.

Finally, finite plasma pressure is known to modify the magnetic configuration in a less controllable manner. As we will discuss in section \ref{SEC_DSJ}, its main effect is to change the average value of $B$ on the flux-surface. While, if $B=B_0$, such modification cannot cause $\delta>0$, if $B=B_0+\delta B_1$, the value of $\delta$ will generally depend on the pressure. $\beta$-dependent deviations from quasi-isodynamicity have been reported e.g. in~\cite{goodman2023qi}. Figure \ref{FIG_DELTA} (right) represents the normalized difference between the magnetic field strength of CIEMAT-QI at $s=0.1$ for $\fsa{\beta}=4$\% and its value in vacuum. A rough estimate of $\delta\sim 10^{-3}$ can be given.

These deviations from perfect neoclassical optimization can be thought of as roughly additive. While the estimates of the previous paragraphs are by no means rigorous bounds, they certainly show the existence of limits to what can be achieved through traditional optimization. Therefore, this approach needs to be complemented with alternative strategies. Since $\delta$ cannot be reduced at will, fields ${B}={B_0}+\delta{B_1}$ with good properties for realistic values of $\delta$ have to be identified. In this work, we will do it by finding ``robust'' QI fields $B_0$. This condition will be precisely formulated in section \ref{SEC_THEORY}.

\section{Orbit-averaged drifts, second adiabatic invariant and transport optimization}\label{SEC_THEORY}


As we have discussed in section \ref{SEC_PICTURE}, in the absence of collisions, particles move at constant $J$. In particular, if $\mathbf{v}_d$ denotes the drift velocity perpendicular to $\mathbf{B}$, and $\mathbf{v}_d\cdot\nabla s$ and $\mathbf{v}_d\cdot\nabla \alpha$ are the radial and tangential drifts, i.e., the components of $\mathbf{v}_d$ that are perpendicular and tangent to the flux-surface, respectively, the following relations exist:
\begin{eqnarray} 
\overline{\mathbf{v}_d\cdot\nabla s} = \frac{m}{\tau_b Z_e\partial_s\Psi}\partial_\alpha J\,,\quad\overline{\mathbf{v}_d\cdot\nabla \alpha} = -\frac{m}{\tau_b Ze\partial_s\Psi}\partial_s J\,.
\end{eqnarray}
Here, $\overline{(...)}=(2/\tau_b)\int_{l_{b_1}}^{l_{b_2}}\mathrm{d}l (...)|v_\parallel|^{-1}$ denotes orbit-average, and $\tau_b\equiv 2\int_{l_{b_1}}^{l_{b_2}}\mathrm{d}l/|v_\parallel|$ is the bounce time. Since the particle drift is generally the sum of the magnetic and the $E\times B$ drift, $\mathbf{v}_d=\mathbf{v}_M+\mathbf{v}_{E}$, it is natural to split
\begin{eqnarray} 
  \partial_s J&=&(\partial_s J)_B +(\partial_s J)_\Phi\,,\quad  (\partial_s J)_B \equiv -\frac{\tau_b Z_e\partial_s\Psi}{m}\overline{\mathbf{v}_M\cdot\nabla \alpha}\,,\quad(\partial_s J)_\Phi \equiv - \frac{\tau_b Z_e\partial_s\Psi}{m}\overline{\mathbf{v}_E\cdot\nabla \alpha}\,,\nonumber\\   
\partial_\alpha J& =&(\partial_\alpha J)_B +(\partial_\alpha J)_\Phi\,,\quad (\partial_\alpha J)_B \equiv\frac{\tau_b Z_e\partial_s\Psi}{m}\overline{\mathbf{v}_M\cdot\nabla s}\,,\quad(\partial_\alpha J)_\Phi \equiv \frac{\tau_b Z_e\partial_s\Psi}{m}\overline{\mathbf{v}_E\cdot\nabla s}\,.   
\end{eqnarray} 

For electrons, fast ions and bulk ions in large aspect ratio stellarators, $(\partial_\alpha J)_\Phi$ (caused by the variation on the flux-surface of the electrostatic potential~\cite{pedrosa2015phi1,estrada2019phi1}) does not play a role in radial transport~\cite{dherbemont2022las} (the situation is different for impurities, see e.g.~\cite{regana2017phi1,calvo2018nf}), and $(\partial_\alpha J)_B=0$ guarantees excellent neoclassical transport properties. For this reason, historically (through reduction of the effective ripple $\varepsilon_{eff}$~\cite{nemov1999neo}), but also in recent works (by explicitly minimizing the distance to perfect quasisymmetry~\cite{landreman2022preciseQS}, to exact quasi-isodynamicity~\cite{jorge2023single,goodman2023qi}, or to other types of perfect omnigeneity~\cite{dudt2023omni}), stellarator optimization has focused on obtaining configurations with very small value of $|(\partial_\alpha J)_B|$ (this quantity is 0 in exactly QS or QI configurations, for which $\varepsilon_{eff}=0$ as well). Nevertheless, it is unclear to what extent very small values of $|(\partial_\alpha J)_B|$ are compatible with other design constraints such as magnetohydrodynamic (MHD) stability, possibility of simple coils or good fast ion confinement at all relevant values of $\beta$. For this reason, in this work we focus on maximizing $-(\partial_s J)_B$, the benefits of which we will discuss in the next paragraphs.

Orbits with $\partial_s J\approx 0$, termed ``superbananas'', have an orbit-averaged drift that is directed in the radial direction and that leads to large radial fluxes. Generally speaking, $|(\partial_\alpha J)_B/(\partial_s J)_B|\sim 1$ is known to cause large fast ion losses ($(\partial_s J)_\Phi$ is negligible for fast ions), and metrics based on reducing this quantity have been proposed~\cite{nemov2008gammac,velasco2021prompt}, although they have only been employed in a few works~\cite{bader2019fastions,leviness2013gammac,sanchez2023qi}. One of the reasons could be that these metrics are based on collisionless physics, and the problem is inherently collisional: fast ions are followed until they are lost or thermalized \textit{via collisions}. Moreover, additional collisionless loss mechanisms besides superbananas exist for fast ions~\cite{paul2022fastions}. For these reasons, direct optimization of guiding center orbits is starting to be proposed as an alternative to QS or QI metrics~\cite{bindel2023direct}.

The effect of the orbit-averaged tangential magnetic drift is also relevant for the bulk transport of optimized stellarators \cite{calvo2017sqrtnu}. Stellarators are usually optimized for a low level of transport in the so-called $1/\nu$ regime \cite{nemov1999neo}. However, the bulk ions of optimized stellarators are expected to be in regimes of lower collisionality, such as the $\sqrt{\nu}$ regime. In QI stellarators, optimization of the bulk energy transport is thus indirect~\cite{beidler2021nature}: minimization of $\varepsilon_{eff}$ (since it encapsulates the configuration dependence of the $1/\nu$ flux) reduces the electron particle flux which, through ambipolarity of the neoclassical fluxes, causes a more negative (with respect to a non-optimized configuration) radial electric field $E_r$. This reduces the ion energy flux, as $Q_i \sim |(\partial_s J)_B + (\partial_s J)_\Phi|^{-3/2}$ \cite{calvo2017sqrtnu}. In a large aspect ratio stellarator ($\epsilon = a/R_0 \ll 1$, where $a$ and $R_0$ are the minor and major radius), $|(\partial_s J)_B|\ll |(\partial_s J)_\Phi|$ typically, which gives the well-known $Q_i \sim E_r^{-3/2}$ scaling. In a tight aspect ratio stellarator, a negative $(\partial_s J)_B$ is expected to improve bulk ion transport for the standard situation of negative $E_r$ and thus negative $(\partial_s J)_\Phi$, specially if $E_r$ is not far from zero. This effect has been largely ignored in stellarator optimization, but we will argue that it can be very important in the core of a reactor. It is also noteworthy that a more negative radial electric field will tend to drive a stronger inward impurity flux. Not relying on a strong negative radial electric field for the bulk energy confinement could also contribute to mitigate the impurity accumulation that is observed in plasmas with reduced turbulent level~\cite{romba2023nbi}.

Finally, a QI field in which $(\partial_s J)_B$ is negative is said to be ``maximum-$J$''. Maximum-$J$ magnetic fields have also been predicted to have a stabilyzing effect on turbulence: the trapped-electron mode (TEM) can be regarded as a drift wave that is driven unstable by a resonance with the precessional drift of trapped electrons; if $(\partial_\alpha J)_B=0$ and $(\partial_s J)_B<0$, the electrons precess on the flux-surface in a direction that contributes to suppress such resonance. For this reason, in maximum-$J$ configurations, density-gradient-driven TEMs are expected to be stable \cite{helander2013tem}. 

In the previous paragraphs, we have argued that a negative $(\partial_s J)_B$ may have a very positive impact on the confinement of energetic and bulk ions, and that this may facilitate optimization, reducing the need for a very small $|(\partial_\alpha J)_B|$. In order to make the argument more explicit, let us assume that the magnetic field strength $B$ of a configuration sufficiently close to QI can be written as in equation (\ref{EQ_SPLITTING}) with $\delta|\nabla B_1|\ll |\nabla B_0|$ (i.e. we consider the case in which $B_1$ has small spatial derivatives). Then, for bulk~\cite{calvo2017sqrtnu,calvo2018jpp} and fast ions, $(\partial_s J)_B$ and $(\partial_\alpha J)_B$ can be expressed to first order in $\delta$ as
\begin{eqnarray}
(\partial_s J)_B&=& \partial_s J_0(s,{\cal E},\mu) + \delta \partial_s J_1(s,\alpha,{\cal E},\mu)\,,\nonumber\\
(\partial_\alpha J)_B &=& \delta \partial_\alpha J_1(s,\alpha,{\cal E},\mu)\,,\label{EQ_SPLITTINGJ}
\end{eqnarray}
where $J_0\sim J_1$. Here, $\partial_\alpha J_1$ is set by both $B_0$ and $B_1$, while $\partial_s J_0$, the dominant term in $(\partial_s J)_B$, is completely determined by $B_0$~\cite{calvo2017sqrtnu,calvo2018jpp}:
\begin{eqnarray}
\partial_\alpha J_1 &=& - \int_{l_{b_{10}}}^{l_{b_{20}}}\frac{2\mu\partial_\alpha B_1}{\sqrt{2({\cal E}-\mu B_0-Ze\Phi_0/m)}}\mathrm{d}l \,,\nonumber\\
\partial_s J_0 &=& - \int_{l_{b_{10}}}^{l_{b_{20}}}\frac{2\mu\partial_s B_0}{\sqrt{2({\cal E}-\mu B_0-Ze\Phi_0/m)}}\mathrm{d}l \,
\end{eqnarray}
where $l_{b_{10}}$ and $l_{b_{20}}$ are the bounce points of trajectories computed along $B_0$ (i.e., ${\cal E}=\mu B_0+Ze\Phi_0/m$ at $l_{b_{10}}$ and $l_{b_{20}}$), $\Phi_0(s)$ is the piece of the electrostatic potential that is constant on the flux-surface. For fast ions, $|Ze\Phi_0/m|\ll\mathcal{E}-\mu B_0$, and the variation of $\Phi$ on the flux-surface has been neglected, as discussed at the beginning of this section.

In this framework, standard stellarator design consists of making $\delta \partial_\alpha J_1$ as small as possible (when explicitly optimizing for QI of QS, by making $\delta$ small). This usually results in poor control over $B_0$ (to give an example, it is not totally uncommon to start an optimization run close to QI and finish it close to QS). In this work, we will focus on targeting the appropriate $B_0$: we will discuss the properties of an optimized field that make $(\partial_s J)_B$ negative \textit{prior to optimization}, and that thus robustly guarantee excellent confinement properties for only moderately small values of $\delta$.

\section{Region of configuration space with robust optimization}\label{SEC_DSJ}

In this section, we discuss the dependence of $\partial_s J_0$ of a QI field on the parameters that describe its magnetic configuration. We will focus on stellarator-symmetric QI fields with a single $B$-\textit{valley} of large aspect ratio stellarators. The relevant parameters are: the $B_{00}(s)$ term in the Fourier expansion of the magnetic field strength $B_0$ in the Boozer angles,
\begin{equation}\label{eq:B_Fourier}
{B_0}(s,\theta,\zeta) = \sum_{n,m} B_{mn}(s)\cos(m\theta-N_{fp}n\zeta)
\end{equation}
(here, $N_{fp}$ is the number of toroidal field periods, the summation runs over $m \ge 0$ and, for $m=0$, over $n\ge 0$), the rotational transform $\iota(s)$, the toroidal and poloidal currents $I_t(s)$ and $I_p(s)$, and the mirror term $B_{M}(s)$. The latter is defined as
\begin{equation}\label{EQ_BM}
B_{M}(s)=B_{max}-B_{00}(s)\,,
\end{equation}
where $B_{max}$ is the maximum of $B_0$ on the flux-surface. In a QI magnetic field, the maxima of $B_0$ lie on a straight line of constant $\zeta$. If we assume, without loss of generality, that $B_0=B_{max}$ at $\zeta=0$, we can write
\begin{equation}
B_{M}(s)=B_0(s,\theta,\zeta=0) - B_{00}(s) = \sum_{n>0} B_{0n}(s)\,.
\end{equation}
Here, we have used that actually $B_0(s,\theta,\zeta=0)$ does not depend on $\theta$; in the Fourier expansion (\ref{eq:B_Fourier}), this implies that $\sum_n B_{mn} = 0$ for $m\neq 0$. 

In appendix~\ref{SEC_CALC}, we show that $\partial_s J_0$ can be approximated by
\begin{eqnarray}
\partial_s J_0 = \hat J
\sqrt{\frac{\mu B_{00}}{{\cal E}}}  
\left[
2 \frac{\partial_s |B_{M}|}{\partial_sB_{00}} E(\kappa^2) 
-
\left(
1+\frac{\partial_s |B_{M}|}{\partial_s B_{00}}
\right)
 K(\kappa^2)
\right]
,
\label{EQ_DSJ0}
\end{eqnarray}
with
\begin{eqnarray}
\hat J= 4\sqrt{{\cal E}/m} 
\frac{I_p + \iota I_t}{N_{fp}(B_{00}|B_{M}|)^{1/2}}
\frac{\partial_s B_{00}}{B_{00}}\,
\label{EQ_HATJ}
\end{eqnarray}
a flux-function that is positive if we assume $\partial_s B_{00}>0$, as in most stellarators. Here 
\begin{equation}
\kappa^2 = \frac{1 - (\mu/{\cal E})\left(B_{00} - |B_{M}|\right)}{2 (\mu/{\cal E}) |B_{M}|}
\end{equation}
takes values between 0 (for deeply trapped ions) and 1 (for barely trapped ions). Equation (\ref{EQ_DSJ0}) is accurate in the limit that we call in this work \textit{standard} mirror. With this expression (in contrast e.g. to \textit{narrow} or \textit{broad} mirror), we refer to a stellarator-symmetric $B_0$ in which trapped particles moving along the field-line that goes through ($\zeta=\pi/N_{fp}$, $\theta=\pi$) encounter a $B_0(\zeta)$ whose variation can be approximately described by a $\cos{(N_{fp}\zeta)}$ term (see appendix~\ref{SEC_B0} for more details). Note that this does not mean that we limit ourselves to quasi-poloidally symmetric fields, as we do not require $B_{m\ne 0,n}$ to be zero. For instance, equation (\ref{EQ_DSJ0}) is valid for exactly QI fields close to the configuration of W7-X.

The value of $\partial_s J_0$ given by equation (\ref{EQ_DSJ0}) is represented in figure~\ref{FIG_DSJ0}, which shows that there is a region of the QI configuration space where $\partial_s J_0<0$ for all trapped particles. This region is determined by $-1 <\partial_s |B_{M}|/\partial_s B_{00}<1$.

\begin{figure}
\begin{center}
\includegraphics[angle=0,width=0.8\columnwidth]{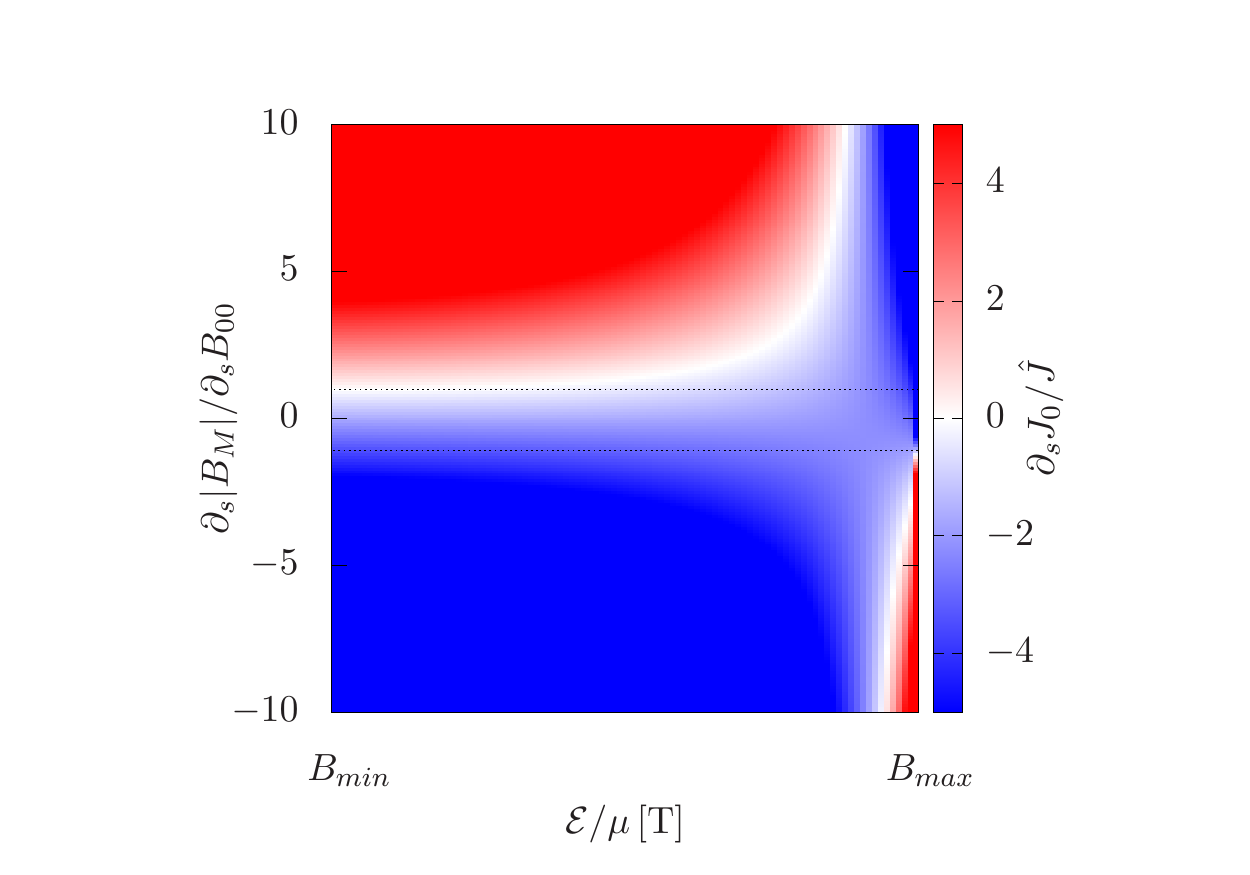}
\end{center}
\caption{Value of $\partial_s J_0$ according to equation (\ref{EQ_DSJ0}) as a function of ${\cal E}/\mu$ and the ratio $\partial_s |B_{M}|/\partial_s B_{00}$. The dotted lines limit the region of the configuration space where $\partial_s J_0<0$ for all particles.}
\label{FIG_DSJ0}
\end{figure}

%

\begin{figure}
\begin{center}
  \begin{minipage}[b]{.33\linewidth}
      {\includegraphics[angle=0,width=1\columnwidth]{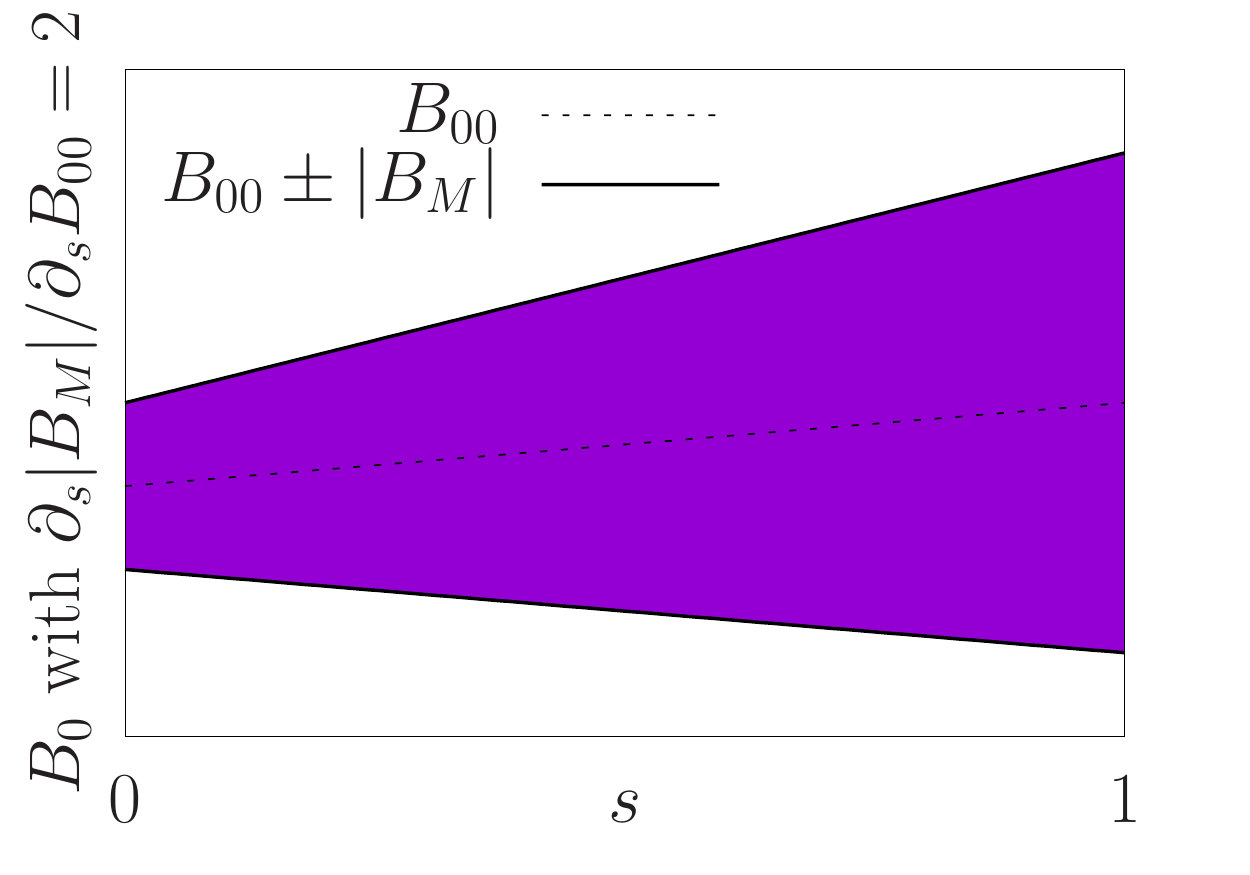}}
      {\includegraphics[angle=0,width=1\columnwidth]{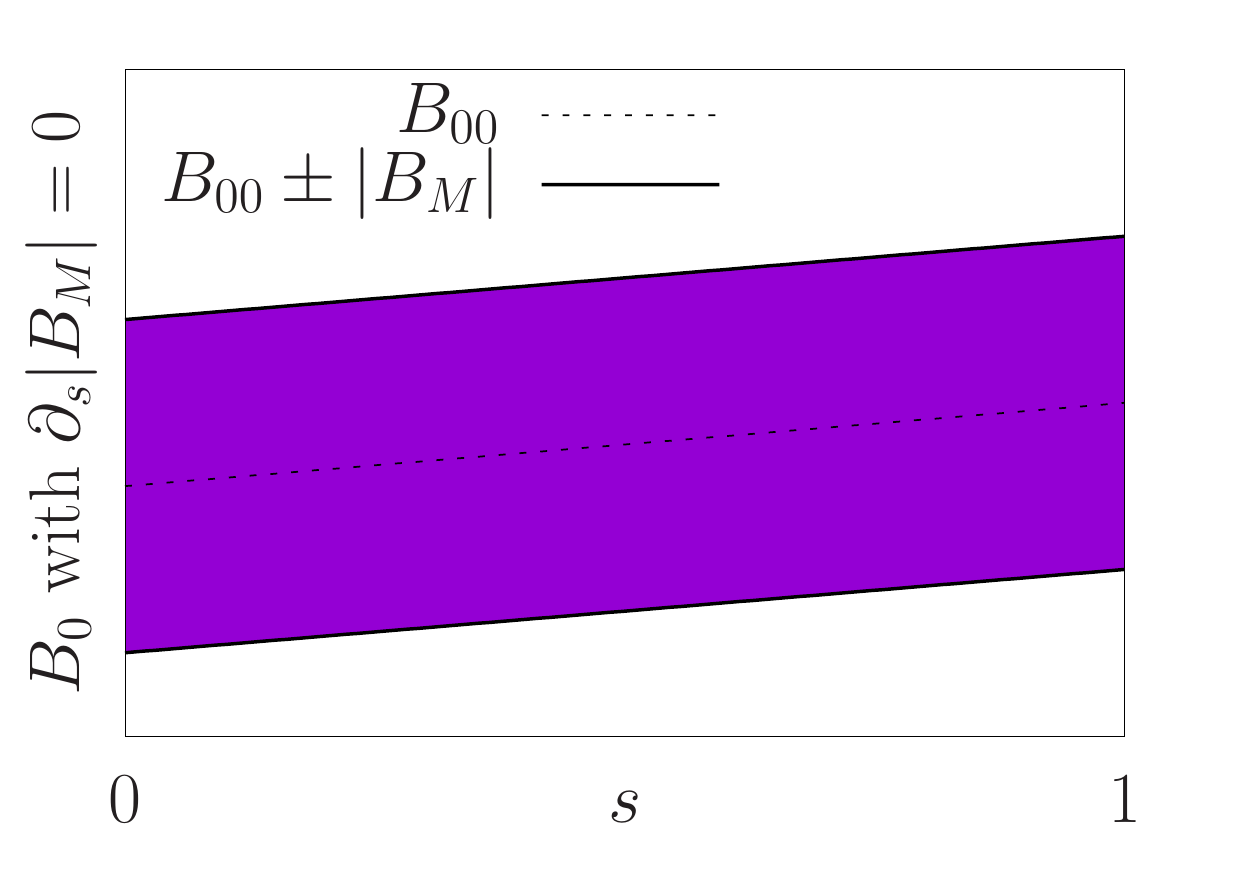}}
  \end{minipage}
\begin{minipage}[t]{.65\linewidth}
          {\includegraphics[angle=0,width=1\columnwidth]{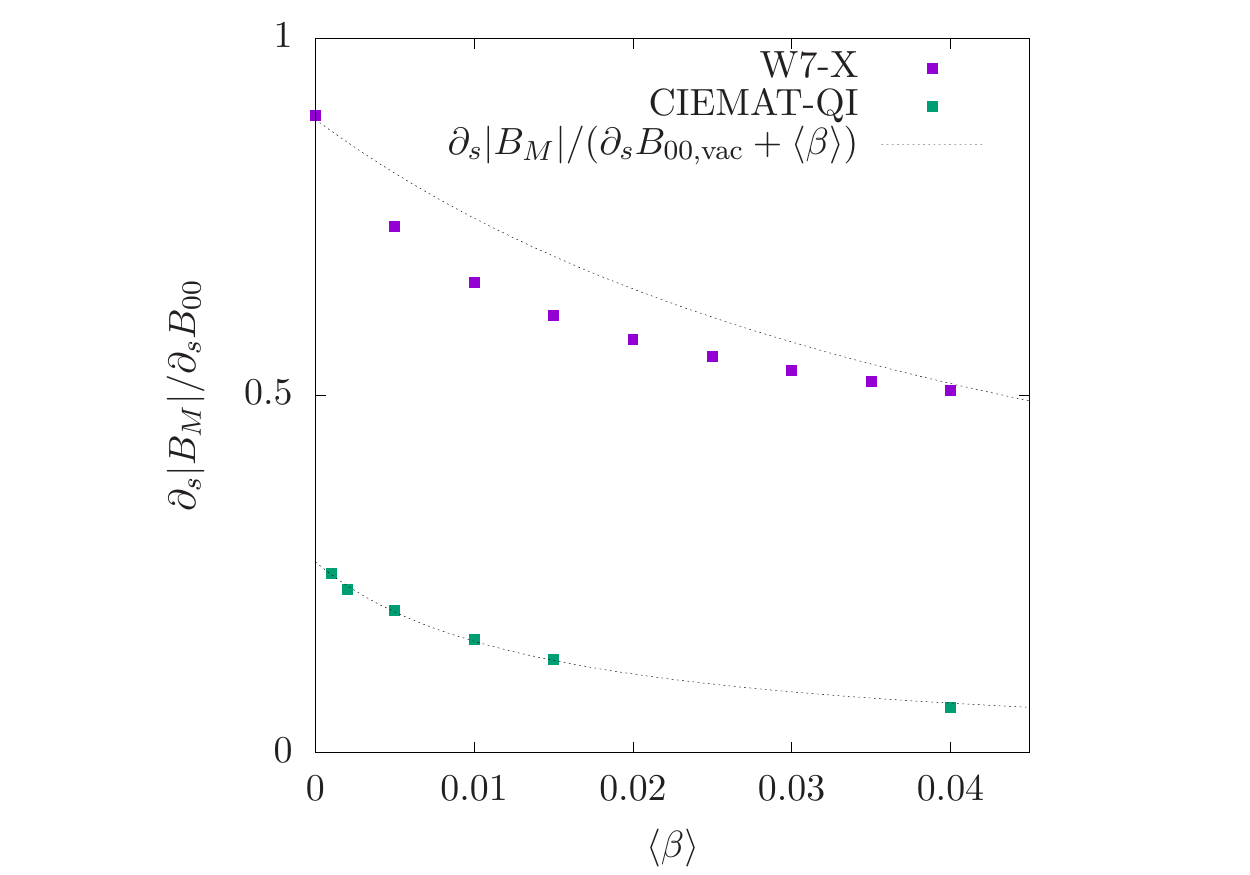}}%
  \end{minipage}%
  \hfill
\end{center}
\caption{Radial variation of $B_0(s,\theta,\zeta)$ for a QI field with $\partial_s|B_M|/\partial_sB_{00}=2$ (left top) and $\partial_s|B_M|/\partial_sB_{00}=0$ (left bottom). For each radial position, $B_{00}-|B_M|\le B_0 \le B_{00}+|B_M|$. Ratio $\partial_s |B_{M}|/\partial_s B_{00}$ for W7-X and CIEMAT-QI as a function of $\fsa{\beta}$ (right).} 
\label{FIG_BM}
\end{figure}

Customarily, $0<\partial_s |B_{M}|/\partial_s B_{00}<1$ has been pursued by resorting to the diamagnetic effect, which makes $\partial_s B_{00}$ larger at finite $\beta$ (here, $\beta = 2\mu_0 p / B^2$, where $p(s)$ is the plasma pressure and $\mu_0$ is the vacuum permeability). In a large aspect ratio stellarator, the effect of finite $\beta \sim\epsilon$ can be absorbed in the $B_{00}(s)$ term,
\begin{equation}
B_{00}(s) \approx  B_{00,\mathrm{vac}}(s)\left(1-\beta(s)/2\right)\,,\label{EQ_BD} 
\end{equation}
see, e.g.,~\cite{dherbemont2022las}. In vacuum, if $\partial_s |B_{M}|/\partial_s|B_{00}|=\partial_s |B_{M}|/\partial_sB_{00,\mathrm{vac}}\gg 1$, superbananas will exist close to the value of $\kappa^2$ such that $2 E(\kappa^2)=K(\kappa^2)$. This means that relative large $\beta$ is needed to remove them. However, if good confinement properties are required at low $\beta$, the design needs to rely on the vacuum magnetic field. In principle, making $\partial_s B_{00,\mathrm{vac}}$ large would also contribute to $\partial_s J_0<0$, but its size is strongly constrained by the MHD equilibrium equations, see e.g.~\cite{dherbemont2022las}. This automatically leaves a small radial derivative of the mirror term $|B_M|$ as the best strategy for negative $\partial_s J_0$. We will say that a magnetic field possesses the flat mirror property if
\begin{equation}
\partial_s |B_M|/\partial_s B_{00}\approx 0\,.\label{EQ_FLATMIRROR}
\end{equation}

Figure \ref{FIG_BM} (left) represents the radial variation of the magnetic field strength in QI fields with and without flat mirror, which allows us to make now the connection between equations (\ref{EQ_DSJ0}) and (\ref{EQ_FLATMIRROR}) and the qualitative discussion of section \ref{SEC_PICTURE}. Figure \ref{FIG_BM} (top left), with $\partial_s|B_M|/\partial_sB_{00}=2$ and for which $\partial_s J_0=0$ for some orbits, corresponds to the QI field sketched in figure \ref{FIG_MIRRORS} (top left), which was not robust against non-omnigeneous perturbations. On the other hand, figure \ref{FIG_BM} (bottom left), with $\partial_s|B_M|=0$ and thus $\partial_s J_0<0$, corresponds to the robust flat-mirror QI field of figure \ref{FIG_MIRRORS} (bottom left). 

It is important to realize that not only the sign, but also the size of $\partial_s J_0$ matters, because the condition $\partial_s J_0<0$ only guarantees absence of superbanana orbits in a perfectly QI stellarator. More negative $\partial_s J_0$ implies that larger $\delta$ can be compatible with the constraint $(\partial_s J)_B=\partial_s J_0+\delta\partial_s J_1<0$. Equation (\ref{EQ_DSJ0}) has thus allowed us to find an avenue for finding perfectly QI fields $B_0(s,\theta,\zeta)$ that fulfill the maximum-$J$ property and, consequently, optimized $B(s,\theta,\zeta)$ fields with robustly good transport properties.

In order to show that the discussion above can help understand qualitatively the confinement properties of a configuration, figure~\ref{FIG_BM} (right) shows the values of $\partial_s|B_{M}|/\partial_s B_{00}$ at $s=0.3$ for two configurations, W7-X (high-mirror) and CIEMAT-QI, as a function of $\fsa{\beta}$) ($\beta$-profiles that are linear in $s$ are assumed throughout this work, so $\partial_s\beta=-2\fsa{\beta}$). We will pay attention to the relative values, as the magnetic fields of these devices are not exactly QI (in this and the next figures, $B_M$ is defined $B_M=\sum_{n>0}B_{0n}$, as in exactly QI fields). W7-X starts from larger values of $\partial_s|B_{M}|/\partial_s B_{00}$ in vacuum, and finite $\beta$ reduces it by half at $\fsa{\beta}=0.04$. On the other hand, CIEMAT-QI has lower values of this ratio already in vacuum. This is in overall agreement with the gradual improvement in the confinement of fast ions expected for W7-X in the range $\fsa{\beta}=0.02 - 0.07$ \cite{velasco2021prompt} and with the good fast ion confinement of CIEMAT-QI at low $\beta$.

It is noteworthy that $\partial_s |B_{M}|/\partial_s B_{00}\ll -1$ causes negative $\partial_s J_0$ for most orbits, with only barely passing particles having $\partial_s J_0>0$. This could result in configurations with good transport properties for processes dominated by deeply trapped particles, as it is the case for fast ions, see section \ref{SEC_FASTIONS}, or if $|(\partial_\alpha J)_B|$ is made very small for the range of $E/\mu$ where $(\partial_s J)_B\approx 0$. This region of the configuration space is left for future studies. Similarly, $\partial_s |B_{M}|/\partial_s B_{00}\gg 1$ could be made (by means of optimization of $(\partial_\alpha J)_B$) consistent with good fast ion confinement, since $\partial_s J_0$ is far from zero in most of the phase space, including deeply trapped particles. However, these configurations would not be expected to have good properties with respect to TEM turbulence, since $\partial_s J_0>0$, i.e., they would be ``minimum-$J$''. Finally, special attention should be paid to the intermediate region in the configuration space around $\partial_s |B_{M}|/\partial_s B_{00}=1$. A configuration designed with $\partial_s |B_{M}|/\partial_s B_{00}>1$ in vacuum may display, as $\beta$ is increased and $\partial_s |B_{M}|/\partial_s B_{00}=1$ is crossed, $\partial_s J_0 \approx 0$ for a wide range of $\mu/E$, and thus large fast ion losses (before entering the region of $\partial_s J_0<0$ as $\beta$ is further increased). Such a non-monotonic behaviour has been observed for instance in~\cite{velasco2021prompt,goodman2023qi}.

We proceed to consider the cases in which approximation (\ref{EQ_HATJ}) does not hold, which we discuss in detail in appendix \ref{SEC_BOTTOM}. QI fields are said to have narrow or broad mirror if the second derivative of $B_0$ with respect to the coordinate along the field line, evaluated at the minimum of $B_0$ on the flux-surface, $B_{min}$, is relatively large or small (a quantitative definition is provided below). When this happens, equation (\ref{EQ_DSJ0}) is not valid, and the calculation of this section changes quantitatively, although not qualitatively. Since, as we will see in section \ref{SEC_FASTIONS} and in appendix~\ref{SEC_BOTTOM}, obtaining exact quasi-isodynamicity is harder for very deeply trapped particles, achieving a very negative $\partial_s J_0$ is more relevant in that region of phase space. In appendix~\ref{SEC_BOTTOM}, we compute $\partial_s J_0$ for deeply trapped orbits in a QI field in which $B_{M}$ is broad or narrow (keeping the requirements of large aspect ratio, stellarator-symmetry, and single $B$-valley). For ${\cal E}/\mu=B_{min}$, one obtains
\begin{eqnarray}
\partial_s J_0 = \hat J
\sqrt{\frac{B_{00}}{B_{min}}}\frac{\pi}{2}\sqrt{\frac{N_{fp}^2|B_M|}{b_w}}
\left(\frac{\partial_s |B_{M}|}{\partial_sB_{00}}-1\right)
\,,
\label{EQ_DSJ0BOTTOMGEN}
\end{eqnarray}
with $b_w=\frac{1}{2}\partial^2_{\zeta}B|_{\alpha=\pi(1-\iota/N_{fp})}$ evaluated at $\zeta=\pi/N_{fp}$, $\theta=\pi$. This is consistent with equation (\ref{EQ_DSJ0}): $\partial_s J_0$ is negative if $\partial_s |B_{M}|/\partial_s B_{00}<1$. If we evaluate equation (\ref{EQ_DSJ0}) at ${\cal E}/\mu=B_{min}$, we obtain
\begin{eqnarray}
\partial_s J_0 = \hat J
\sqrt{\frac{B_{00}}{B_{min}}}\frac{\pi}{2}
\left(\frac{\partial_s |B_{M}|}{\partial_sB_{00}}-1\right)
\,.
\label{EQ_DSJ0BOTTOM}
\end{eqnarray}
Comparison of equations (\ref{EQ_DSJ0BOTTOMGEN}) and (\ref{EQ_DSJ0BOTTOM}) shows that a broad mirror ($b_w<N_{fp }^2|B_{M}|$) will make $\partial_s J_0$ more negative and a narrow mirror ($b_w>N_{fp }^2|B_{M}|$) will make it closer to zero. This may be one of the factors that have led to the good fast ion confinement of configurations discussed in \cite{drevlak2014fastions,drevlak2018rose} and of CIEMAT-QI. A result equivalent to equation (\ref{EQ_DSJ0BOTTOMGEN}) has been recently obtained in \cite{rodriguez2023maxj}.

We end this section by mentioning that equation (\ref{EQ_DSJ0}) can be generalized for any omnigeneous magnetic field, i.e., not necessarily quasi-isodynamic. For a general omnigeneous field, in which the maxima of $B_0$ lie on straight lines of constant $M\theta - N_{fp}N\zeta$, the result for $\partial_sJ_0$ is
\begin{eqnarray}
  \partial_s J_0 = \hat J \frac{N_{fp}}{|N_{fp}N-M\iota|}
\sqrt{\frac{\mu B_{00}}{{\cal E}}}  
\left[
2 \frac{\partial_s |B_{M}|}{\partial_sB_{00}} E(\kappa^2) 
-
\left(
1+\frac{\partial_s |B_{M}|}{\partial_s B_{00}}
\right)
 K(\kappa^2)
\right]
,
\label{EQ_DSJ0omni}
\end{eqnarray}
where now $B_M$ can be computed (assuming that the maxima of $B_0$ lie on $M\theta - N_{fp}N\zeta = 0$)
\begin{equation}
  B_{M}(s)=B_0(s,\theta,\zeta=M\theta/(N_{fp}N)) - B_{00}(s) = \sum_{k>0} B_{kM,kN}(s)\,.\label{EQ_BMomni}
\end{equation}
Inspection of equations (\ref{EQ_DSJ0omni}) and (\ref{EQ_BMomni}) shows that the main difference with respect to the QI case is that, for non-QI omnigeneous fields (i.e., for omnigenous fields with helicities different than $M=0$), $\partial_s |B_{M}|$ is positive and relatively large, and $\partial_s |B_{M}|/\partial_s B_{00}$ is generally further from zero. This means that $\partial_s J_0>0$ for most particles, with $\partial_s J_0=0$ at a value of $\kappa^2$ such that $2 E(\kappa^2)\approx K(\kappa^2)$, close to the boundary between passing and trapped, and weakly dependent on $\beta$. The optimization strategy should then be different than for a QI field (the standard approach would be making $|\partial_\alpha J|$ very small at least in some region of phase space) and we leave it for future studies.

\begin{figure}
\begin{center}
\includegraphics[angle=0,width=0.49\columnwidth]{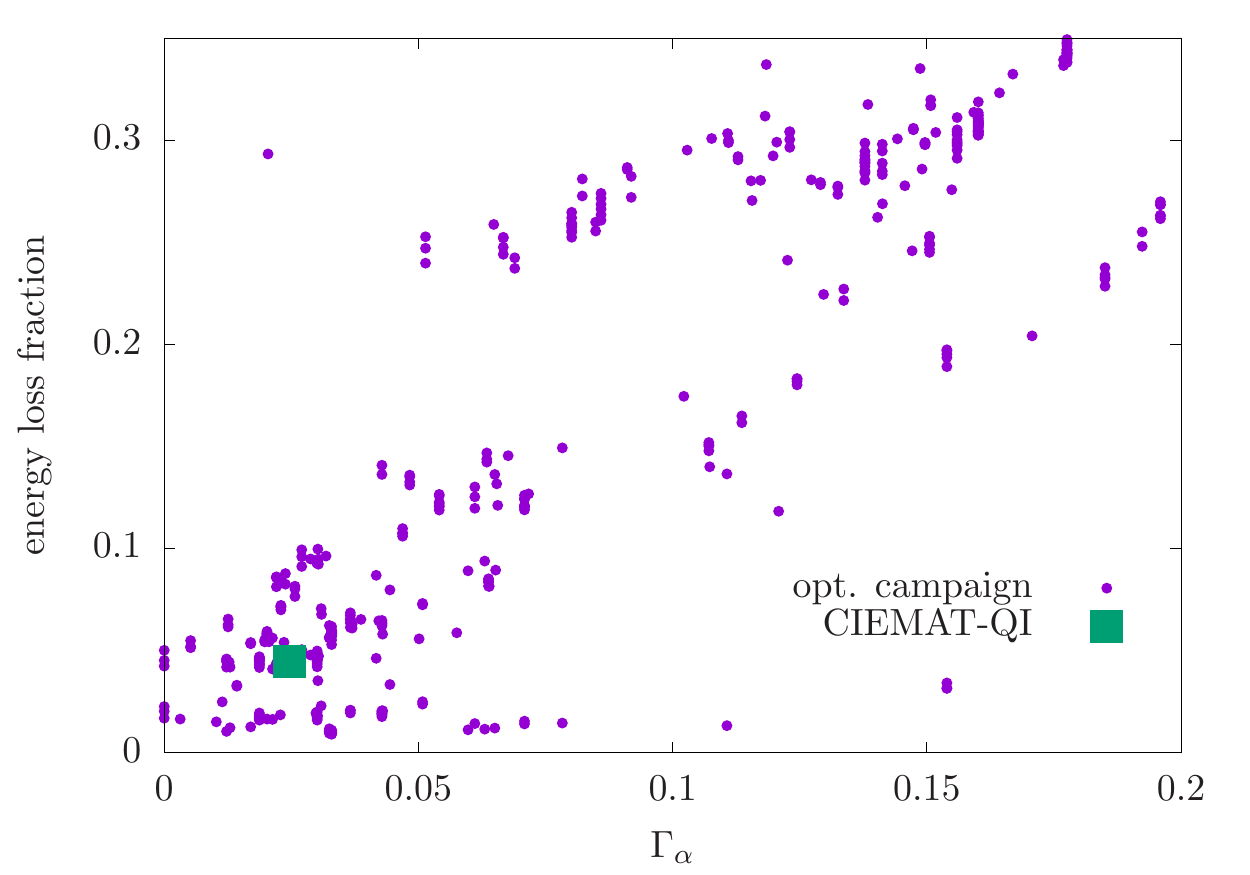}
\includegraphics[angle=0,width=0.49\columnwidth]{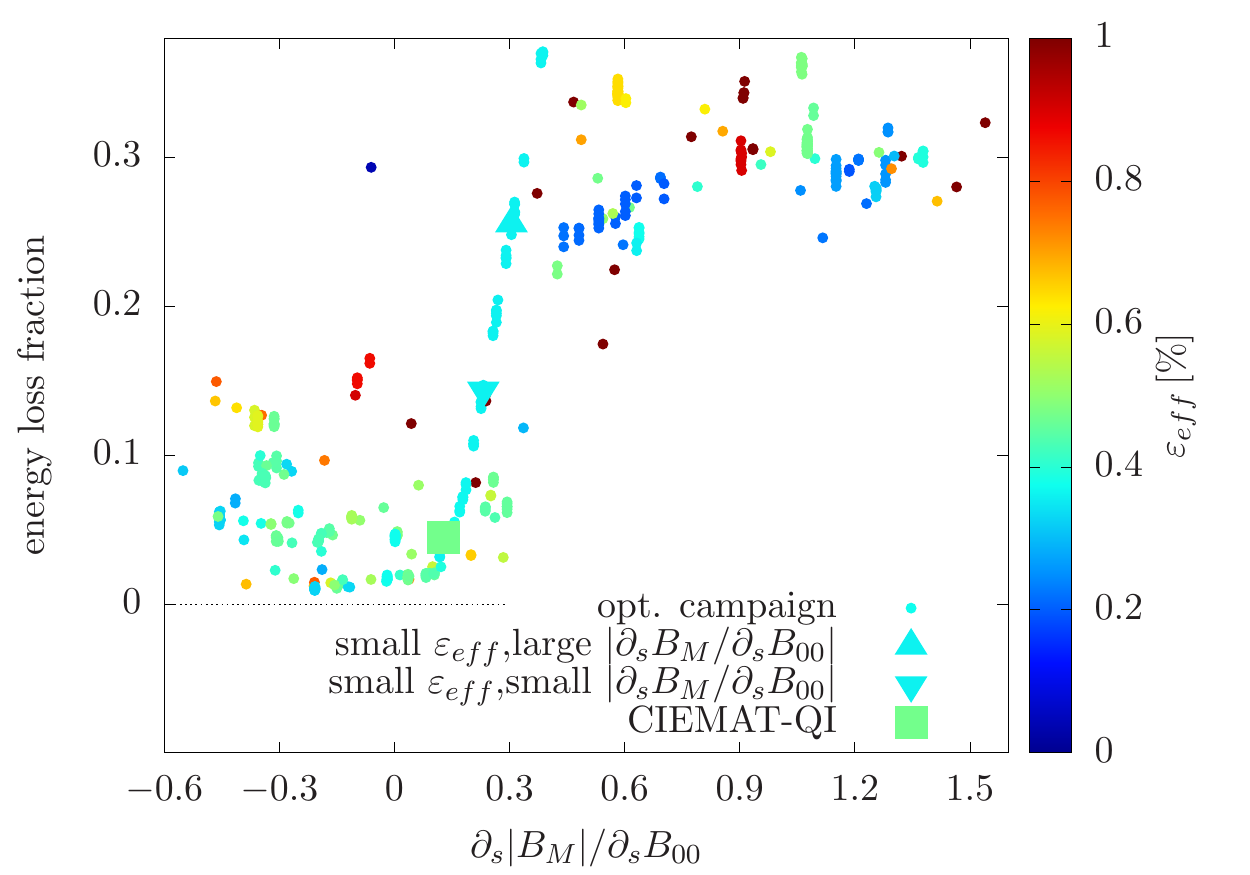}
\end{center}
\caption{Energy loss fraction versus $\Gamma_\alpha$ (left) and $\partial_s|B_M|/\partial_s B_{00}$ (right) at $s=0.3$.}
\label{FIG_LOSSPROXY}
\end{figure}

\section{Fast ion confinement}\label{SEC_FASTIONS}

We start this section by evaluating the fast ion confinement of a selection of the configurations that were explored during the optimization campaign that led to CIEMAT-QI~\cite{sanchez2023qi}. We first scale them to reactor size by setting their minor radius and $B$ on axis to be those of ARIES-CS~\cite{najmabadi2008ariescs} (the original value of $\fsa{\beta}$, between 0 and 0.05, is preserved). Then, we distribute 3.5$\,$MeV alpha particles uniformly in velocity space and in the flux-surface, with a radial distribution that represents fusion-born alpha particles, as in~\cite{landreman2022bootstrap}. Then, we employ the code \texttt{ASCOT} to follow their guiding-center trajectories in the presence of collisions until they either thermalize or go through the last closed flux-surface. In order to confirm that our study is addressing the right physics, figure~\ref{FIG_LOSSPROXY} (left) represents the fraction of energy that is lost versus $\Gamma_\alpha$~\cite{velasco2021prompt} at $s=0.3$ (all the flux-surface quantities of this section are evaluated at this flux-surface, which encloses the volume where most alpha particles are born). $\Gamma_\alpha$, similarly to $\Gamma_c$~\cite{nemov2008gammac}, is a fast-particle transport proxy that relies on the size of $(\partial_\alpha J)_B/(\partial_s J)_B$, and the good correlation between its values and the energy losses means that the loss mechanism that we discuss in this work (orbit-averaged trajectories that go through the last closed flux-surface) is dominant even in the presence of collisions. Figure~\ref{FIG_LOSSPROXY} (right) tries to assess whether it is the minimization of $(\partial_\alpha J)_B$ or the maximization of $|(\partial_s J)_B|$ that has led to the good properties of CIEMAT-QI, and to clarify the role of $\partial_s|B_{M}|/\partial_s B_{00}$. The most salient feature is that, at nearly constant $\varepsilon_{eff}$ (around 0.4\%-0.5\%), a threshold-like behaviour is observed with the loss fraction quickly falling from 25\% to around 5\% when $\partial_s|B_{M}|/\partial_s B_{00}$ is decreased. Then, the loss fraction stays roughly constant around $\partial_s|B_{M}|/\partial_s B_{00} \approx 0$ and starts to increase again when $\partial_s|B_{M}|/\partial_s B_{00}$ becomes negative enough. Configurations with values of $\varepsilon_{eff}$ as large as 1\% and small $\partial_s|B_{M}|/\partial_s B_{00}$ have fast ion confinement properties similar to configurations with larger $\partial_s|B_{M}|/\partial_s B_{00}$ and much smaller $\varepsilon_{eff}$ (and the configurations with smallest ripple, $\sim$0.1\%, are the ones with largest losses).

In order to discuss the main features of figure~\ref{FIG_LOSSPROXY} (right), figure~\ref{FIG_LOSSLAMBDA} (left) represents $(\partial_s J)_B$ versus ${\cal E}/\mu$ for three configurations that have been highlighted in figure~\ref{FIG_LOSSPROXY}: two of them have similarly low $\varepsilon_{eff}$ and different values of $\partial_s|B_{M}|/\partial_s B_{00}$ (labelled 'large' and 'small' in the figure) that lead to very different energy losses. The other one is CIEMAT-QI (at $\fsa{\beta}=0.015$). The main qualitative difference between the configurations is the extent of the region of phase-space where $(\partial_s J)_B\approx 0$. For CIEMAT-QI it is very small, leading to very low prompt losses (the ones that should be associated to large values os $|(\partial_\alpha J)_B/(\partial_s J)_B|$ and that channel most of the energy lost). 

\begin{figure}
\begin{center}
\includegraphics[angle=0,width=0.49\columnwidth]{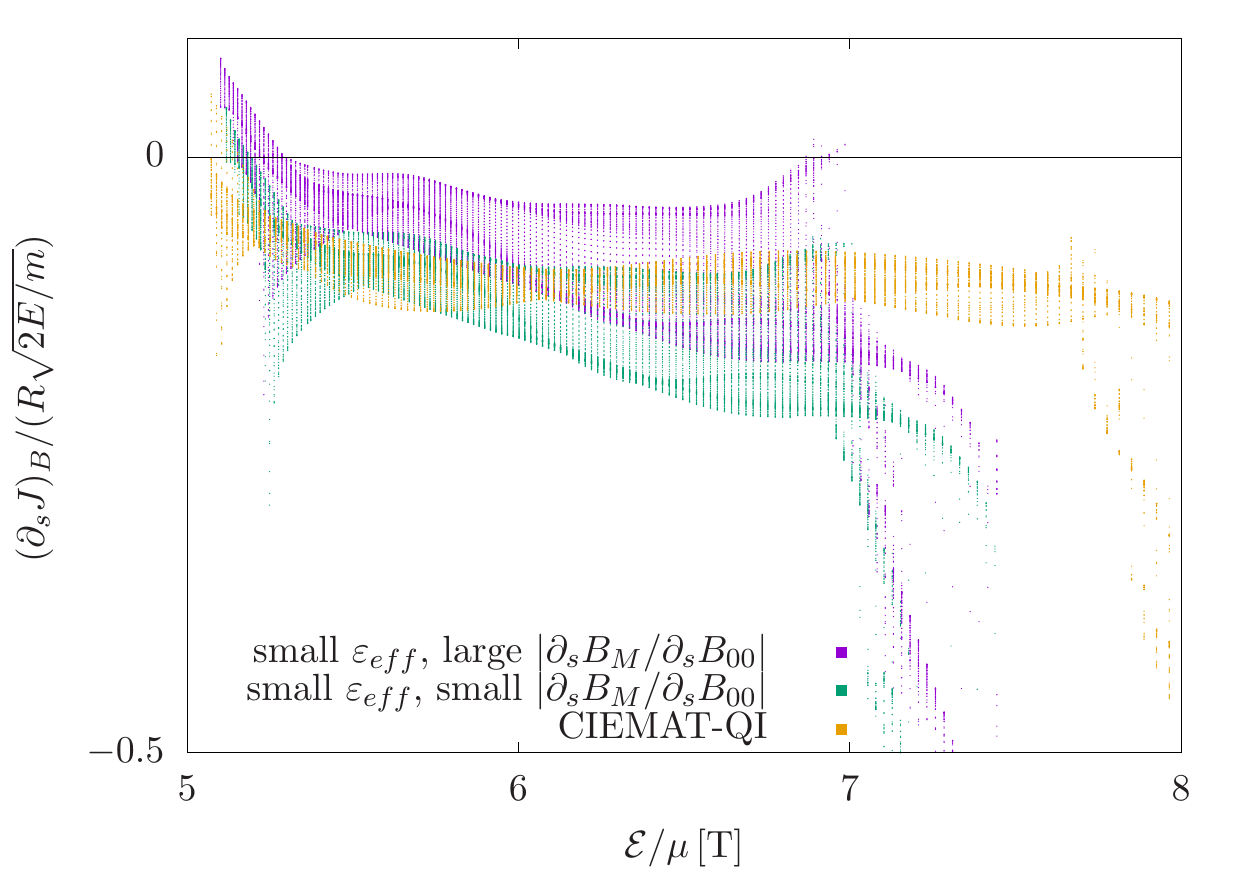}
\includegraphics[angle=0,width=0.49\columnwidth]{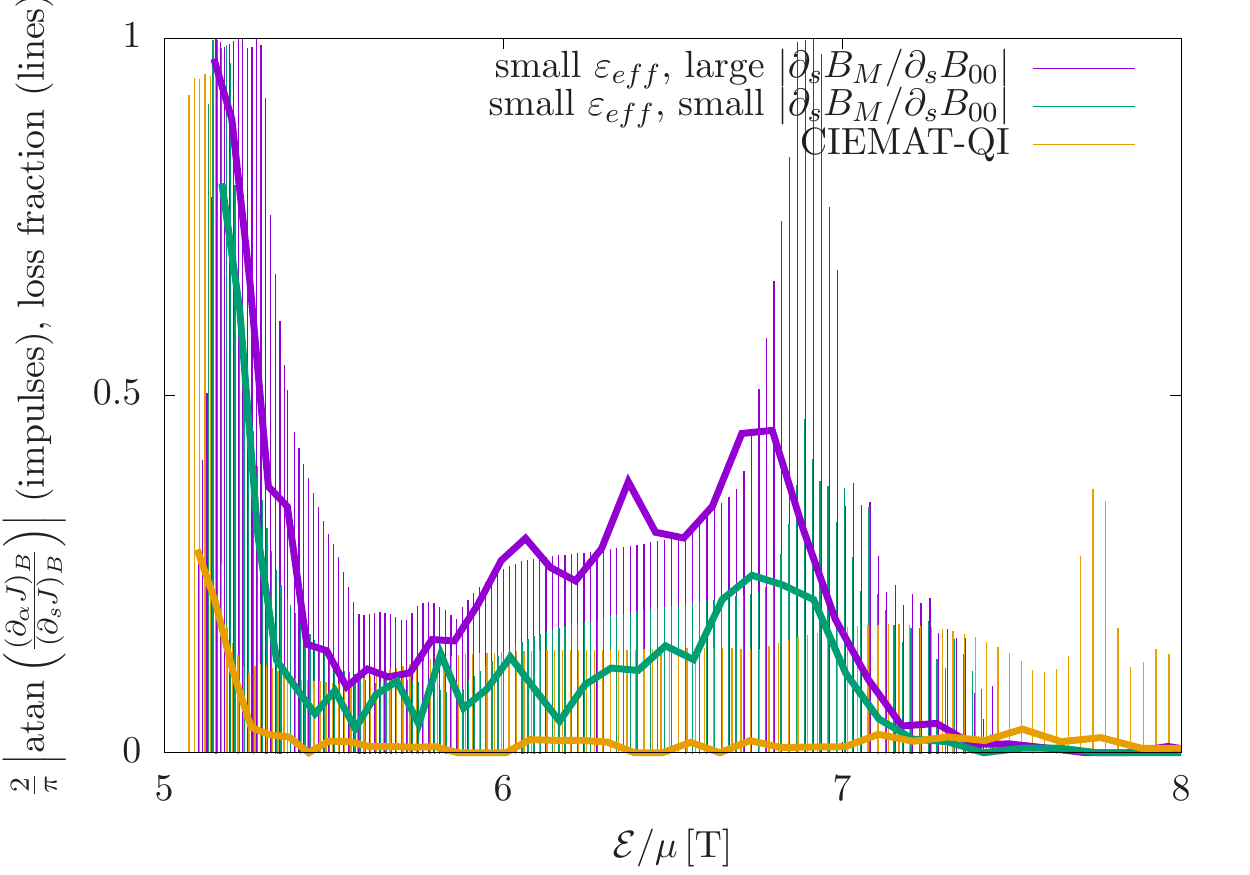}
\end{center}
\caption{$(\partial_s J)_B$ (left) and size of $(\partial_\alpha J)_B/(\partial_s J)_B$ ratio at $s=0.3$ compared with the fraction of prompt losses (right) for selected configurations as a function of ${\cal E}/\mu$.}
\label{FIG_LOSSLAMBDA}
\end{figure}

In order to complete the picture, figure~\ref{FIG_LOSSLAMBDA} (right) represents $|(\partial_\alpha J)_B/(\partial_s J)_B|$ (reescaled so that it takes values between 0 and 1) as a function of ${\cal E}/\mu$ together with the prompt losses. The two configurations with the same $\varepsilon_{eff}$, that actually come from a "$\beta$-scan" performed by the optimizer between $\fsa{\beta}=0.01$ and $\fsa{\beta}=0.02$, have indistinguisheable values of $(\partial_\alpha J)_B$ that, combined with different $(\partial_s J)_B$ lead to values that correlate remarkably well with the fast ion losses (which  peak close to where $(\partial_s J)_B\approx 0$). For CIEMAT-QI, $|(\partial_\alpha J)_B/(\partial_s J)_B|$ is below 0.2 for most of the velocity space, a value that has been associated with very low losses~\cite{velasco2021prompt}, as indeed checked with \texttt{ASCOT}.

We note that, for each value of $\mathcal{E}/\mu$ in figure~\ref{FIG_LOSSPROXY} there are several values of $(\partial_s J)_B$ and $(\partial_\alpha J)_B/(\partial_s J)_B$. They correspond to different field lines, as both quantities are functions of $\alpha$ for $\delta>0$, see equation (\ref{EQ_SPLITTINGJ}). The wider range of values for $\mathcal{E}/\mu\approx B_{min}$ indicate that exact quasi-isodynamicity is harder to obtain for deeply trapped orbits, as we will explain in appendix~\ref{SEC_BOTTOM}.

The threshold behaviour observed in~\ref{FIG_LOSSPROXY} (right) can now be interpreted as follows: reducing $\partial_s|B_{M}|/\partial_s B_{00}$ makes $\partial_s J_0$ more negative and, for a given size of $(\partial_\alpha J)_B$, it reduces the region of small $|(\partial_s J)_B|$. Once $\partial_s|B_{M}|/\partial_s B_{00}$ is close enough to zero, no further gains can be expected after $(\partial_s J)_B$ becomes negative for all trapped particles. The loss fraction may saturate at a level different than zero if mechanisms different than superbananas play a (minor) role on fast ion transport.

\section{Bulk ion transport}\label{SEC_BULKIONS}

For bulk ions, it can be demonstrated~\cite{calvo2018jpp} that $(\partial_s J)_B \sim \epsilon(\partial_s J)_\Phi$ if $E_r=- \partial_r\Phi_0\sim T_D/(a Z_De)$ (in this section, the subscripts $D$, $T$ and $e$ refer to deuterium and tritium ions and to electrons respectively, and we will use the radial coordinate $r=a\sqrt{s}$). For this reason (and because doing so makes the calculation easier), the effect of $(\partial_s J)_B$ is almost always neglected in large aspect ratio stellarators. However, while $E_r\sim T_D/(a Z_De)$ is a standard assumption if the plasma is in the so-called ion root \cite{dinklage2013ncval} and the ambipolarity equation is dominated by the ions, it is inaccurate in low collisionality plasmas with $T_e\approx T_D$, where $E_r$ can be closer to zero, see e.g. \cite{velasco2017hole}. This is the situation in the core of a fusion-grade plasma and it is found to be ubiquitous in predictive transport simulations of low-collisionality plasmas in concurrence with a flat ion temperature profile~\cite{turkin2011predictive,geiger2014w7x,sunnpedersen2015op11,warmer2016burning,warmer2018itg,beidler2018density}. It is straightforward to show that this ($\partial_s T_D\approx 0$ and $E_r\approx 0$) is a stable solution of the energy transport equation: following~\cite{beidler2021nature,velasco2017hole}, we can write the neoclassical particle and energy fluxes
\begin{eqnarray}
 \frac{\Gamma_b}{n_b}&=&- L_{11}^b\left[\left(\frac{\partial_r n_b}{n_b}- \frac{Z_beE_r}{T_b}\right)+\delta_{12}^b\frac{\partial_r T_b}{T_b}\right]\,,\label{EQ_FLUX}\\
 \frac{Q_b}{n_bT_b}&=&- L_{11}^b\left[\delta_{21}^b\left(\frac{\partial_r n_b}{n_b}- \frac{Z_beE_r}{T_b}\right)+\delta_{22}^b\frac{\partial_r T_b}{T_b}\right]\,,\label{EQ_EFLUX}
\end{eqnarray}
where $L_{11}^b$ and $L_{11}^b\delta^b_{ij}$ are neoclassical thermal transport coefficients (with their usual definition, see e.g.~\cite{beidler2021nature}), obtained by solving the drift-kinetic equation of species $b$. We will assume a fusion plasma with $T_e=T_D=T_T$, $n_e=2n_D=2n_T$ and $\partial_r n_b= 0$ (in the calculations of ~\cite{turkin2011predictive,geiger2014w7x,sunnpedersen2015op11,warmer2016burning,warmer2018itg,beidler2018density}, relatively flat density profiles are assumed, as observed in large experimental stellarators with core heating and external fuelling, due to particle transport being dominated by turbulent diffusion~\cite{thienpondt2023density}). The steady state profiles of the temperature and radial electric field are given by the solution of the energy balance and ambipolarity equations
\begin{eqnarray}
\sum_b Z_be\Gamma_b= 0\,,\label{EQ_AMB}\\
Q_b=\frac{1}{r}\int_0^r\mathrm{d}r' r' P_b\,,\label{EQ_EBAL}
\end{eqnarray}
where $P_b$ is the net energy source to species $b$. It is illutrative to rearrange equations (\ref{EQ_AMB}) and (\ref{EQ_EBAL}) into
\begin{eqnarray}
  E_r&=& \left(\frac{\partial_r T_D}{e}\right)\left[\frac{\delta_{12}^D L_{11}^D +\delta_{12}^T L_{11}^T-2\delta_{12}^e L_{11}^e} {L_{11}^D +L_{11}^T+2L_{11}^e}\right]\,,\label{EQ_AMB2}\\
 \frac{\partial_r T_D}{T_D}&=& -(L_{11}^D)^{-1}\left[\delta_{22}^D-\delta_{21}^D\left(\frac{\delta_{12}^D L_{11}^D +\delta_{12}^T L_{11}^T-2\delta_{12}^e L_{11}^e} {L_{11}^D +L_{11}^T+2L_{11}^e}\right)  \right]^{-1}\frac{1}{r}\int_0^r\mathrm{d}r' r' P_D\,.\label{EQ_EBAL2}
\end{eqnarray}
Expressions  (\ref{EQ_AMB2}) and (\ref{EQ_EBAL2}) are equations, since $L_{11}^b$ and $\delta^b_{ij}$ generally depend on $E_r$. Neverthless, for most fusion-relevant scenarios (see e.g.~\cite{beidler2021nature}), the combinations of the transport coefficients within the square brackets are positive $\mathcal{O}(1-10)$ quantities, with a weak dependence on $E_r$~\cite{beidler2021nature}. When this is the case, flat ion temperature will cause $E_r$ to be close to zero, according to equation  (\ref{EQ_AMB2}). This will make $L_{11}^D$ large, which will keep the ion temperature gradient small, according to equation (\ref{EQ_EBAL2}). In summary, a small $E_r$ and hence a small $(\partial_s J)_\Phi$, can be very deleterious for the performance of a reactor, which would need more heating in order to access the desired temperatures close to the axis.
\begin{figure}
\begin{center}
\includegraphics[angle=0,width=0.49\columnwidth]{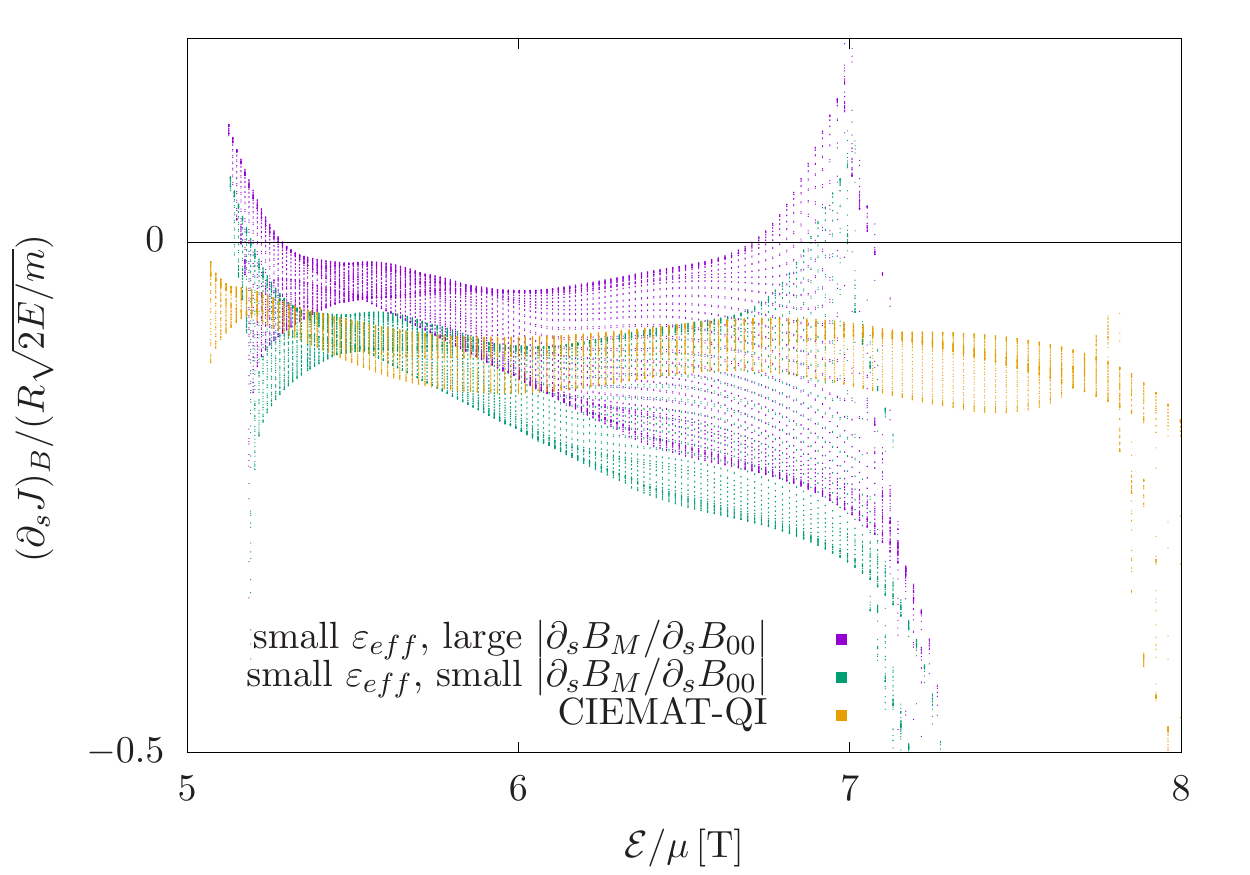}
\includegraphics[angle=0,width=0.49\columnwidth]{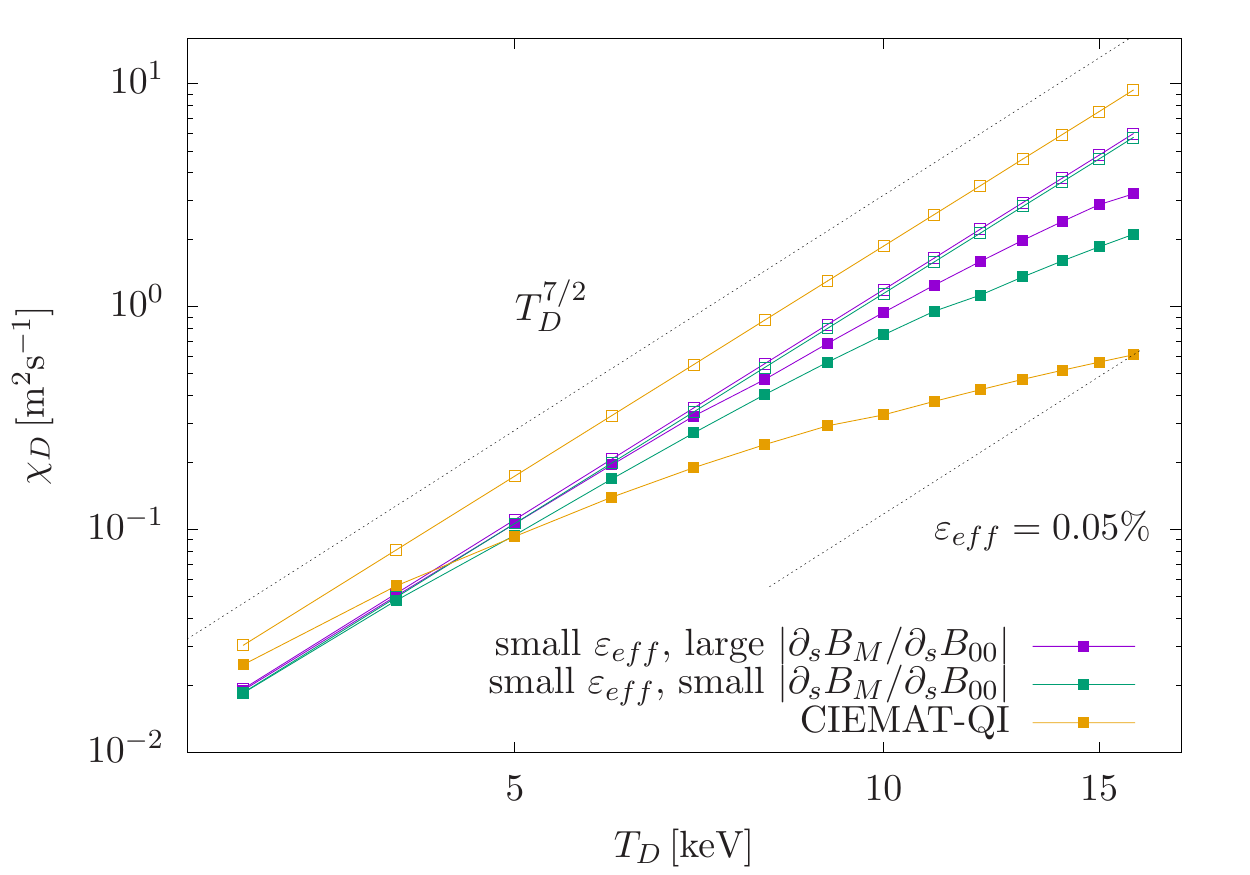}
\end{center}
\caption{$(\partial_s J)_B$ as a function of ${\cal E}/\mu$ (left) and deuterium heat diffusivity versus temperature (right) at $s=0.1$. For the heat diffusivity, full squares represent the correct calculation and open squares the calculation with $(\partial_s J)_B$ artificially set to zero.}
\label{FIG_TISCAN}
\end{figure}

A negative $(\partial_s J)_B$ could help reduce the neoclassical transport coefficients and thus alleviate this problem. If $E_r$ is close to zero and $(\partial_s J)_B \approx 0$, radially-local neoclassical calculations will predict an extremely large $1/\nu$ transport (in reality, the radial magnetic drift should also be included in the calculation \cite{satake2006fortec3d}, but this means that the bulk ion transport would be radially global, and thus unacceptably large for a reactor, as would also be the case for fast ions losses). Finite $(\partial_s J)_B$ contributes to reduce the neoclassical flux in these cases.  Figure \ref{FIG_TISCAN} (left) shows $(\partial_s J)_B$ at $s=0.1$ as a function of ${\cal E}/\mu$ for the three configurations of section \ref{SEC_FASTIONS} (all the flux-surface quantities of this section are evaluated at this flux-surface, in the region where flat ion temperature is sometimes predicted). In configurations close to omnigeneity, for which equations (\ref{EQ_SPLITTINGJ}) are valid, the bulk ions will typically be in the superbanana-plateau regime \cite{calvo2017sqrtnu}, since $(\partial_s J)_B$ will usually be zero in some region of the phase-space. This may be expected to be the case of the two configurations with reduced $\varepsilon_{eff}$ of figure \ref{FIG_TISCAN} (left). Conversely, if $(\partial_s J)_B$ has no zeros, the ions are predicted to be in the $\sqrt{\nu}$ regime (a $\sqrt{\nu}$ regime in which $Q_i \sim |(\partial_s J)_B|^{-3/2}$ \cite{calvo2017sqrtnu}, instead of the standard  $Q_i \sim |(\partial_s J)_\Phi|^{-3/2}$). This should be case of CIEMAT-QI.

Figure \ref{FIG_TISCAN} (right) shows with full symbols the deuterium heat diffusivity  $\chi_D\equiv L_{11}^D\delta_{22}^D$ versus the ion temperature. During the scan, the following quantities are kept constant: $n_e=2n_D=2n_T=2\times 10^{20}\,$m$^{-3}$, $T_e=T_D=T_T$, $E_r=0$. The calculation is done with \texttt{KNOSOS} \cite{velasco2021knosos}. For most of the temperature scan, the two configurations with smaller rippple ($\varepsilon_{eff}\approx 0.24\%$) show the expected $T_D^{7/2}$-dependence of the $1/\nu$ regime. Only at very high temperature the effect of the tangential magnetic drift starts to show. This happens earlier for the configuration with smaller $\partial_s|B_{M}|/\partial_s B_{00}$, whose $(\partial_s J)_B$ is more negative. The effect of the tangential magnetic drift is much stronger for CIEMAT-QI which, already from moderate temperatures, is transitioning towards the $T_D^{-1/4}$ scaling of a $\sqrt{\nu}$ regime associated to the tangential magnetic drift~\cite{calvo2017sqrtnu}. The overall result is a neoclassical transport that is one order of magnitude smaller despite having a larger effective ripple. Although $\varepsilon_{eff}\approx 0.33\%$ for CIEMAT-QI at $s=0.1$, the configuration has (at $T_D=15\,$keV) the level of transport for $E_r=0$ of a stellarator with $(\partial_s J)_B=0$ and $\varepsilon_{eff}\approx 0.05\%$. This is the same order of magnitude of the lowest values of effective ripple ever reported for a nearly QI design~\cite{goodman2023qi} and, to our knowledge, below any value obtained at reactor $\beta$-values. The heat diffusivity is around 0.6\,m$^2$s$^{-1}$ even for $E_r=0$. This is close to the value $\chi_D/\chi_{GB}\sim 3\times 10^{-2}$ predicted to be compatible with stellarator reactor physics design points in~\cite{alonso2022reactor} (here, the gyro-Bohm diffusivity is defined as $\chi_{GB}=v_{T,D}^3\Omega_D^{-2}a^{-1}$, with $v_{T,D}$ the thermal velocity of deuterium and $\Omega_D$ its gyrofrequency).


For reference, the open symbols in figure \ref{FIG_TISCAN} (right) represent calculations in which $(\partial_s J)_B$ is artificially set to zero, as standard neoclassical codes do \cite{beidler2011ICNTS}. This highlights the relevance for reactor scenarios of solving a drift-kinetic equation that includes the tangential magnetic drift in the ion orbits. Accurate predictive transport simulations~\cite{banonnavarro2023tango} are also important, since they can be employed to asses reactor scenarios. For instance, it can be argued that, in many situations, the existence of a radial electric field of standard size will render the differences of figure \ref{FIG_TISCAN} (right) smaller. Nevertheless, the fact that many predictive simulations~\cite{turkin2011predictive,geiger2014w7x,sunnpedersen2015op11,warmer2016burning,warmer2018itg,beidler2018density} do display scenarios of $E_r$ very close to zero make advisable to devise optimization strategies for the bulk energy transport that do not rely completely on $E_r$, i.e., that go beyond the minimization of the $1/\nu$ flux. This has been the case of CIEMAT-QI.

Figure \ref{FIG_CHIPROXY} shows the deuterium heat diffusivity for $T_D=12\,$keV at $s=0.1$ as a function of $\Gamma_c$ (left) and versus $\partial_s|B_{M}|/\partial_s B_{00}$ (right). The two figures resemble those of figure \ref{FIG_LOSSPROXY}: there is good correlation between the heat diffusivity and the size of $(\partial_\alpha J)_B/(\partial_s J)_B$, and robustly small values of $(\partial_\alpha J)_B/(\partial_s J)_B$ can be achieved by making $\partial_s|B_{M}|/\partial_s B_{00}$ decrease below a certain threshold value.

\begin{figure}
\begin{center}
\includegraphics[angle=0,width=0.49\columnwidth]{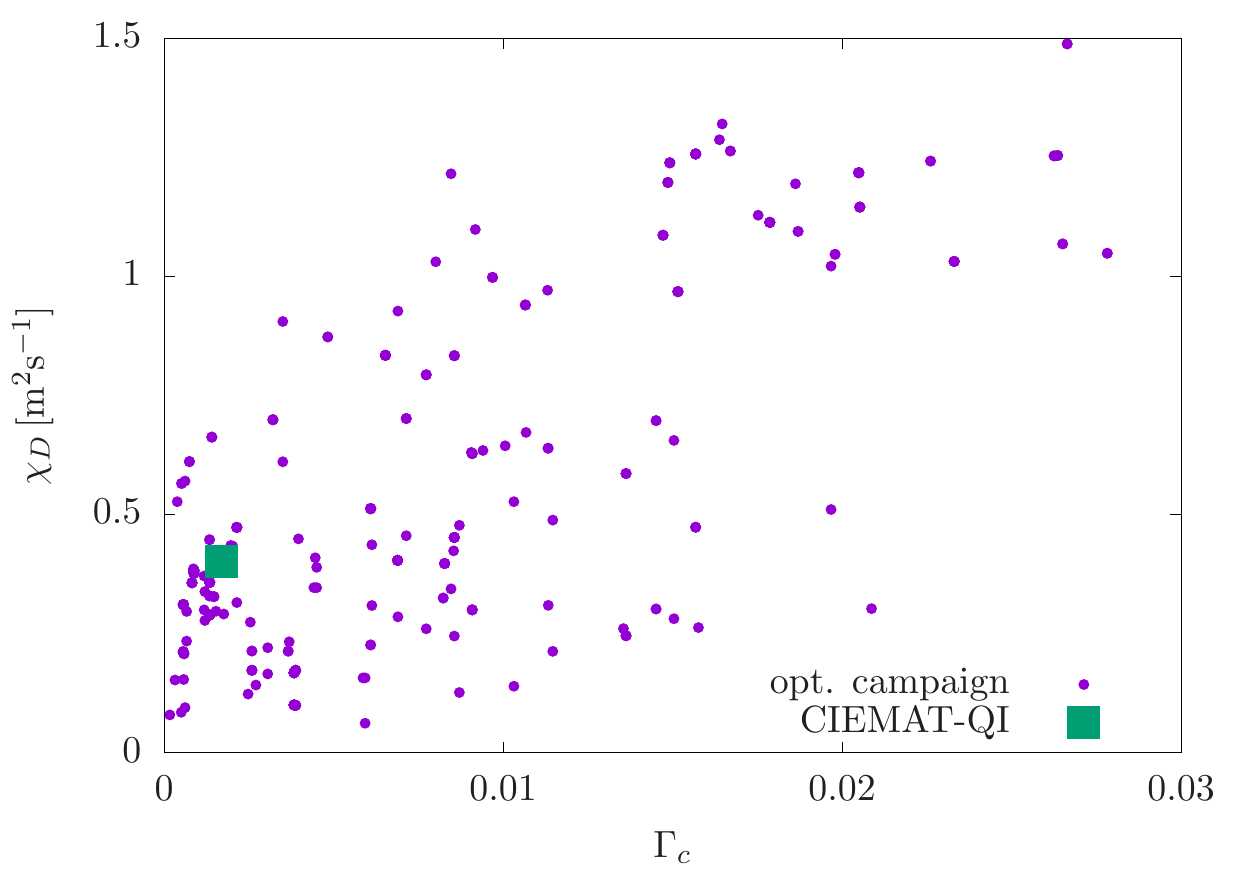}
\includegraphics[angle=0,width=0.49\columnwidth]{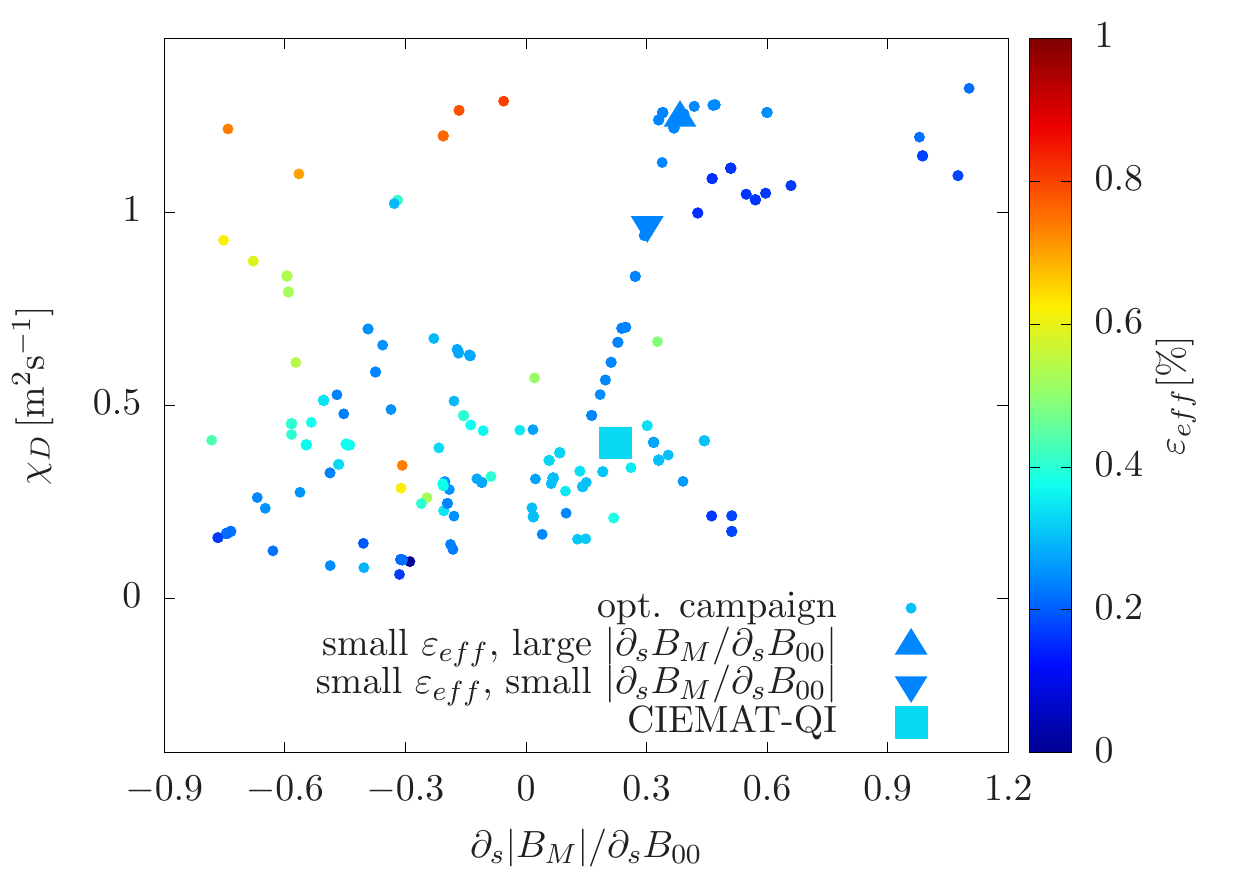}
\end{center}
\caption{Deuterium heat diffusivity versus $\Gamma_c$ (left) and $\partial_s|B_M|/\partial_s B_{00}$ (right).}
\label{FIG_CHIPROXY}
\end{figure}

Finally, we note that the effect of $-(\partial_s J)_B$ on bulk transport may have consequences for the impurities in these scenarios of flat temperature and radial electric field relatively close to zero. In order to illustrate this, the radial electric field of equation (\ref{EQ_AMB2}) can be input into equation (\ref{EQ_FLUX}) in order to compute the particle flux of a trace light impurity. Imposing $\Gamma_I=0$ gives the steady-state impurity density profile in the absence of impurity sources
\begin{equation}
 \frac{\partial_r n_I}{n_I}= \left[Z_I\left(\frac{\delta_{12}^D L_{11}^D +\delta_{12}^T L_{11}^T-2\delta_{12}^e L_{11}^e} {L_{11}^D +L_{11}^T+2L_{11}^e}\right)-\delta_{12}^I\right]\frac{\partial_r T_D}{T_D}\approx  \left[Z_I\delta_{12}^D-\delta_{12}^I\right]\frac{\partial_r T_D}{T_D}\,,\label{EQ_ACC}
\end{equation}
where it has been employed that $|L_{11}^e|\ll |L_{11}^D| \sim |L_{11}^T|$ if the radial electric field is close enough to zero. If the bulk ions are roughly in the $1/\nu$ regime (as it is the case of the configuration of small $\partial_s |B_M|/\partial B_{00}|$ in figure \ref{FIG_TISCAN} (right)), $\delta_{12}^D=\delta_{12}^T\approx 7/2$, and equation~(\ref{EQ_ACC}) predicts impurity accumulation ($\delta_{12}^I\approx 7/2$ and $\delta_{12}^I\approx 3/2$ if the impurity is in the $1/\nu$ or plateau regime, respectively). If, thanks to a finite $(\partial_s J)_B$, the bulk ions are in the superbanana-plateau regime, $\delta_{12}^D=\delta_{12}^T\approx 1$, and screening of helium is possible; if they are in the small-$E_r$ $\sqrt{\nu}$ regime, $\delta_{12}^D=\delta_{12}^T\approx -1/4$, and the impurities tend to be expelled (because $E_r$ becomes slightly positive, see equation (\ref{EQ_AMB2})). Generally speaking, if the bulk ions are in a regime of mild $T_D$-dependence, as it is the case of CIEMAT-QI in figure \ref{FIG_TISCAN} (right), screening of light impurities should be possible (and the finiteness of $\delta_{12}^e L_{11}^e$ could also contribute in this respect~\cite{velasco2017hole}). This will happen because a small radial electric field can coexist with some peaking of the ion temperature. It should be noted that accumulation or expulsion of impurities will ultimately depend on the size of ${\partial_r T_D/T_D}$ combined with the relative importance of the classical and turbulent transport channels, which provide non-negligible convective terms~\cite{buller2019cla,regana2021imp}.

\section{Discussion}\label{SEC_DISCUSSION}

In this work, we have identified flat-mirror quasi-isodynamic fields as a \textit{family} of configurations with robustly good transport properties, thanks to maximization of $-(\partial_s J)_B$ in vacuum}
. They show improved confinement of bulk and fast ions with respect to configurations that are closer to perfect omnigeneity, and for a wider range of plasma parameters. 

The potential benefits of a flat mirror are not limited to the neoclassical transport of energetic and bulk ions. If, thanks to a non-negligible $-(\partial_s J)_B$, a very small radial electric field can coexist with some peaking of the ion temperature, temperature screening of light impurities and helium ashes may be possible in some reactor scenarios. Moreover, although in this work we have focused on the neoclassical properties, a magnetic field that has the maximum-$J$ property has also been predicted to have a stabilyzing effect on TEM turbulence. Future work should address more in detail, through gyrokinetic simulations, the question whether targeting for flat mirror magnetic fields may be also a potential route for the reduction of the turbulent heat fluxes of ions and electrons at low $\beta$.

The magnetic configuration CIEMAT-QI~\cite{sanchez2023qi} belongs to this family of robustly-optimized stellarators. It is, already at low $\beta$, close to fulfilling the maximum-$J$ property. As a result of this, with an effective ripple comparable to W7-X, it presents much better fast and bulk ion confinement, and these properties are largely maintained when its magnetic configuration is generated by filamentary coils. Work is ongoing to design other QI configurations, based on the same concept, with different number of field periods~\cite{godino2023qi}.


The results presented in this work have potential benefits for a future stellarator reactor design. Because flat-mirror QI configurations do not need to be very close to exact omnigeneity, it should be easier to find buildable coils using two-stage optimization, and the resulting configuration should be more robust against error fields. Additionally, because they do not rely on high $\beta$ or a large radial electric field in order to confine the fast and bulk ions, respectively, it should be easier to safely reach and maintain the reactor scenario. Better confinement of bulk and fast ions imply less auxiliary heating and lower risk of wall damage by the fast ions. Both things are a concern not only once the reactor reaches its physics design point, but also on earlier phases of the reactor discharge.

 \section*{Acknowledgements}

 This work has been carried out within the framework of the EUROfusion Consortium and has received funding from the Euratom research and training programme 2014-2018 and 2019-2020 under grant agreement No. 633053. The views and opinions expressed herein do not necessarily reflect those of the European Commission. This research was supported in part by grant PID2021-123175NB-I00, Ministerio de Ciencia, Innovaci\'on y Universidades, Spain. This work was supported by the U.S. Department of Energy under contract number DE-AC02-09CH11466. The United States Government retains a non-exclusive, paid-up, irrevocable, world-wide license to publish or reproduce the published form of this manuscript, or allow others to do so, for United States Government purposes. The authors had enriching discussions about fast ion optimization of W7-X-like configurations with C.D. Beidler and M. Drevlak. Some of the ideas of this paper were developed for/during the the Simons Collaboration on Hidden Symmetries and Fusion Energy Annual Meeting, March 2023. The authors had fruitful discussions with Arturo Alonso, Daniel Carralero, Javier Escoto, Jos\'e Manuel Garc\'ia-Regaña, Guillermo Godino-Sedano, Eduardo Rodr\'iguez and Per Helander.

\appendix

\section{Calculation of $\partial_s J$}\label{SEC_CALC}

We will use a right-handed system of spatial coordinates $\{s, \theta, \zeta\}$ in which $\theta$ and $\zeta$ are the Boozer angles. Then, the magnetic field can be written as
\begin{equation}\label{eq:B_contravariant}
\bB = \frac{\partial_s\Psi}{\sqrt{g}} \left(\partial_\zeta \bx + \iota \partial_\theta \bx\right),
\end{equation}
and also as
\begin{equation}\label{eq:B_covariant}
\bB = I_t \nabla\theta + I_p \nabla\zeta + \omega \nabla s.
\end{equation}
Here, $\bx(s,\theta,\zeta)$ gives the position in Euclidean space,
\begin{equation}
\sqrt{g} (s,\theta,\zeta) = \frac{\partial_s\Psi}{\bB\cdot\nabla\zeta} = (\partial_s \bx \times \partial_\theta \bx) \cdot \partial_\zeta \bx
\end{equation}
is the volume element, $I_t(s)$ and $I_p(s)$ are the toroidal and poloidal currents, and $\omega(s,\theta,\zeta)$ is such that $\bB\cdot\nabla s = 0$. Taking the dot product of equations \eq{eq:B_contravariant} and \eq{eq:B_covariant}, one finds that
\begin{equation}\label{eq:Jacobian_Boozer_coor}
\frac{1}{\bun\cdot\nabla\zeta} = \frac{I_p + \iota I_t}{B} \, .
\end{equation}

Since we are only interested in $(\partial_s J)_B$, we will use as velocity variables $v$, the magnitude of the velocity, and $\lambda=\mu/{\cal E}$, the pitch angle coordinate, both of which are conserved in the absence of a radial electric field. In these variables, the second adiabatic invariant is defined for trapped trajectories by
\begin{equation}\label{eq:J}
J(s,\alpha,v,\lambda) = 2 v \int_{\zeta_{b_1}}^{\zeta_{b_2}}\sqrt{1-\lambda B} \,
\frac{1}{\bun\cdot\nabla\zeta} \, 
\dd\zeta,
\end{equation}
where $\bun = B^{-1}\bB$, $\zeta_{b_1}$ and $\zeta_{b_2}$ are the bounce points of the trajectory, and the integral is taken keeping $\theta - \iota\zeta$ constant.

Differentiating equation \eq{eq:J} with respect to $s$, noting that the integrand vanishes at the bounce points, and using equation \eq{eq:Jacobian_Boozer_coor},
\begin{equation}\label{EQ_DIF1}
\partial_s \left( \frac{1}{\bun\cdot\nabla\zeta} \right)
=
\frac{\partial_s(I_p + \iota I_t)}{B} - \frac{I_p + \iota I_t}{B^2}\partial_s B|_\alpha
\end{equation}
and
\begin{equation}\label{EQ_DIF2}
\partial_s \sqrt{1-\lambda B}|_\alpha
=
\frac{-\lambda \partial_s B|_\alpha}{2\sqrt{1-\lambda B}},
\end{equation}
we obtain
\hskip-2.5cm\begin{equation}\label{eq:d_sJ}
(\partial_s J)_B = 2 v \int_{\zeta_{b_1}}^{\zeta_{b_2}}
\Bigg[
\sqrt{1-\lambda B}\Bigg(
\frac{\partial_s(I_p + \iota I_t)}{B} - \frac{I_p + \iota I_t}{B^2}\partial_s B|_\alpha
\Bigg)
-
\frac{\lambda \partial_s B|_\alpha}{\sqrt{1-\lambda B}}\,
\frac{I_p + \iota I_t}{2B}
\Bigg]
\dd\zeta.
\end{equation}
In equations (\ref{EQ_DIF1}), (\ref{EQ_DIF2}) and (\ref{eq:d_sJ}), we have emphasized that the derivative of $B$ is taken at constant $\alpha$.

Equation~(\ref{eq:d_sJ}) is valid for a generic stellarator. We will now particularize it for a large aspect ratio quasi-isodynamic field. Any stellarator-symmetric QI field, $B=B_0$, with one magnetic \textit{valley} can be written~\cite{cary1997omni,parra2015omni} as
\begin{equation}\label{eq:B_ETA}
{B_0}(s,\theta,\zeta) = B_{00}(s) + B_{M}(s) \cos(N_{fp}\eta)\,
\end{equation}
(where stellarator-symmetry, periodicity and quasi-isodynamicity constrain the possible functions $\eta(\theta,\zeta)$). For the sake of simplicity, in this appendix we perform the calculation for $N_{fp}=1$, and below we provide the general result. We assume that $B_{M}$ has constant sign, at least in a neighborhood of the value of $s$ where we are calculating. For the moment, we take $B_{M} <0$ (if $B_{M} > 0$, as in equation (\ref{eq:B_Fourier}), one can make a change of integration variable $\zeta\mapsto \pi - \zeta$ in equation \eq{eq:d_sJ} and then follow almost identical steps). Then,
\begin{eqnarray}\label{eq:aux_expression}
\sqrt{1-\lambda B_0} = \sqrt{1 - \lambda(B_{00} + B_{M} \cos\eta)} =
\nonumber\\[5pt]
\hspace{1cm}
\sqrt{1 - \lambda(B_{00} - B_{M} + B_{M} (1 + \cos\eta))}=
\nonumber\\[5pt]
\hspace{1cm}
\sqrt{1 - \lambda(B_{00} - B_{M} + 2B_{M}\cos^2(\eta/2))}=
\nonumber\\[5pt]
\hspace{1cm}
\sqrt{1 - \lambda(B_{00} + B_{M} - 2B_{M}\sin^2(\eta/2))}=
\nonumber\\[5pt]
\hspace{1cm}
\sqrt{2 \lambda |B_{M}|}
\,
\sqrt{
\kappa^2 - \sin^2(\eta/2)
}
\,
,
\end{eqnarray}
where
\begin{equation}
\kappa^2 = \frac{1 - \lambda(B_{00} + B_{M})}{2 \lambda |B_{M}|} \in [0,1].
\end{equation}
Using equation \eq{eq:aux_expression} and keeping only the lowest order terms in $\epsilon \ll 1$ in equation \eq{eq:d_sJ}, we get
\begin{eqnarray}\label{eq:d_sJ_2}
\partial_s J_0 &=& 
2 v
\sqrt{2 \lambda |B_{M}|}
\,
\frac{\partial_s(I_p + \iota I_t)}{B_{00}} 
 \int_{\eta_{b_1}}^{\eta_{b_2}}
\sqrt{
\kappa^2 - \sin^2(\eta/2)
}
\,
\frac{\partial\zeta}{\partial\eta}\dd\eta
-
 v 
\frac{I_p + \iota I_t}{B_{00}}
\sqrt{\frac{\lambda}{2 |B_{M}|}}
\\[5pt]
\hspace{0.5cm}
&\times&
\left[
\partial_s B_{00}
\int_{\eta_{b_1}}^{\eta_{b_2}}
\frac{1}{
\,
\sqrt{
\kappa^2 - \sin^2(\eta/2)
}}\,
\frac{\partial\zeta}{\partial\eta}\dd\eta
+
\partial_s B_{M}
\int_{\eta_{b_1}}^{\eta_{b_2}}
\frac{ \cos\eta}{
\,
\sqrt{
\kappa^2 - \sin^2(\eta/2)
}}\,
\frac{\partial\zeta}{\partial\eta}\dd\eta
\right]
.
\nonumber
\end{eqnarray}
The first term on the right-hand side of equation \eq{eq:d_sJ_2} is typically small, but we keep it because there could be particular cases in which its effect can be noticeable. In order to obtain a simple expression for $\partial_s J_0$, we need to make the additional assumption, that we will justify in appendix~\ref{SEC_B0}, that
\begin{equation}\label{EQ_VALIDITY}
\left|\frac{\partial\zeta}{\partial\eta}-1\right|\ll 1\,
\end{equation}
along the integration domain. When this is the case, employing
\begin{eqnarray}
\int_{\eta_{b_1}}^{\eta_{b_2}}
\sqrt{
\kappa^2 - \sin^2(\eta/2)
}\,
\dd\eta
=
4 E(\kappa^2)
+
4(\kappa^2 -1) K(\kappa^2),
\end{eqnarray}
\begin{eqnarray}
\int_{\eta_{b_1}}^{\eta_{b_2}}
\frac{1}{\sqrt{
\kappa^2 - \sin^2(\eta/2)
}}
\,
\dd\eta
=
4 K(\kappa^2)
\end{eqnarray}
and
\begin{eqnarray}
\int_{\eta_{b_1}}^{\eta_{b_2}}
\frac{ \cos\eta}{
\,
\sqrt{
\kappa^2 - \sin^2(\eta/2)
}}\,
\dd\eta
=
8 E(\kappa^2) - 4 K(\kappa^2),
\end{eqnarray}
where
\begin{eqnarray}
K(\kappa^2)
=
\int_0^{\pi/2}
\frac{1}{\sqrt{1-\kappa^2\sin^2 x}} \dd x
\end{eqnarray}
is the complete elliptic integral of the first kind and
\begin{eqnarray}
E(\kappa^2)
=
\int_0^{\pi/2}
\sqrt{1-\kappa^2\sin^2 x} \, \dd x
\end{eqnarray}
is the complete elliptic integral of the second kind, we can rewrite equation \eq{eq:d_sJ_2} as
\begin{eqnarray}\label{eq:d_sJ_3}
\partial_s J_0 = 
8 v
\frac{\partial_s(I_p + \iota I_t)}{B_{00}} 
\sqrt{2 \lambda |B_{M}|}
\,
 \Bigg(
E(\kappa^2)
+
(\kappa^2 -1) K(\kappa^2)
\Bigg)
\\[5pt]
\hspace{1cm}
-
 4v 
\frac{I_p + \iota I_t}{B_{00}}
\sqrt{\frac{\lambda}{2 |B_{M}|}}
\,
\partial_s B_{M}
\left[
2 E(\kappa^2) 
+
\left(
\frac{\partial_s B_{00}}{\partial_s B_{M}} -1
\right)
 K(\kappa^2)
\right]
.
\nonumber
\end{eqnarray}
Expression \eq{eq:d_sJ_3} is valid for $B_M < 0$ and $N_{fp}=1$. It can be checked that the result for any value of $N_{fp}$ and arbitrary sign of $B_M$ is
\begin{eqnarray}\label{eq:d_sJ_4}
\partial_s J_0 = 
8 v
\frac{\partial_s(I_p + \iota I_t)}{N_{fp}B_{00}}
\sqrt{2 \lambda |B_{M}|}
\,
 \Bigg(
E(\kappa^2)
+
(\kappa^2 -1) K(\kappa^2)
\Bigg)
\\[5pt]
\hspace{1cm}
+
 4v 
\frac{I_p + \iota I_t}{N_{fp}B_{00}}
\sqrt{\frac{\lambda}{2 |B_{M}|}}
\,
\partial_s|B_{M}|
\left[
2 E(\kappa^2) 
-
\left(
\frac{\partial_s B_{00}}{\partial_s|B_{M}|} +1
\right)
 K(\kappa^2)
\right]
.
\nonumber
\end{eqnarray}
When the current through the plasma is small, as it is usually the case in QI stellarators, the first term on the right hand side of equation (\ref{eq:d_sJ_4}) can be neglected.

\begin{figure}[ht]
\begin{center}
\includegraphics[angle=0,width=0.47\columnwidth]{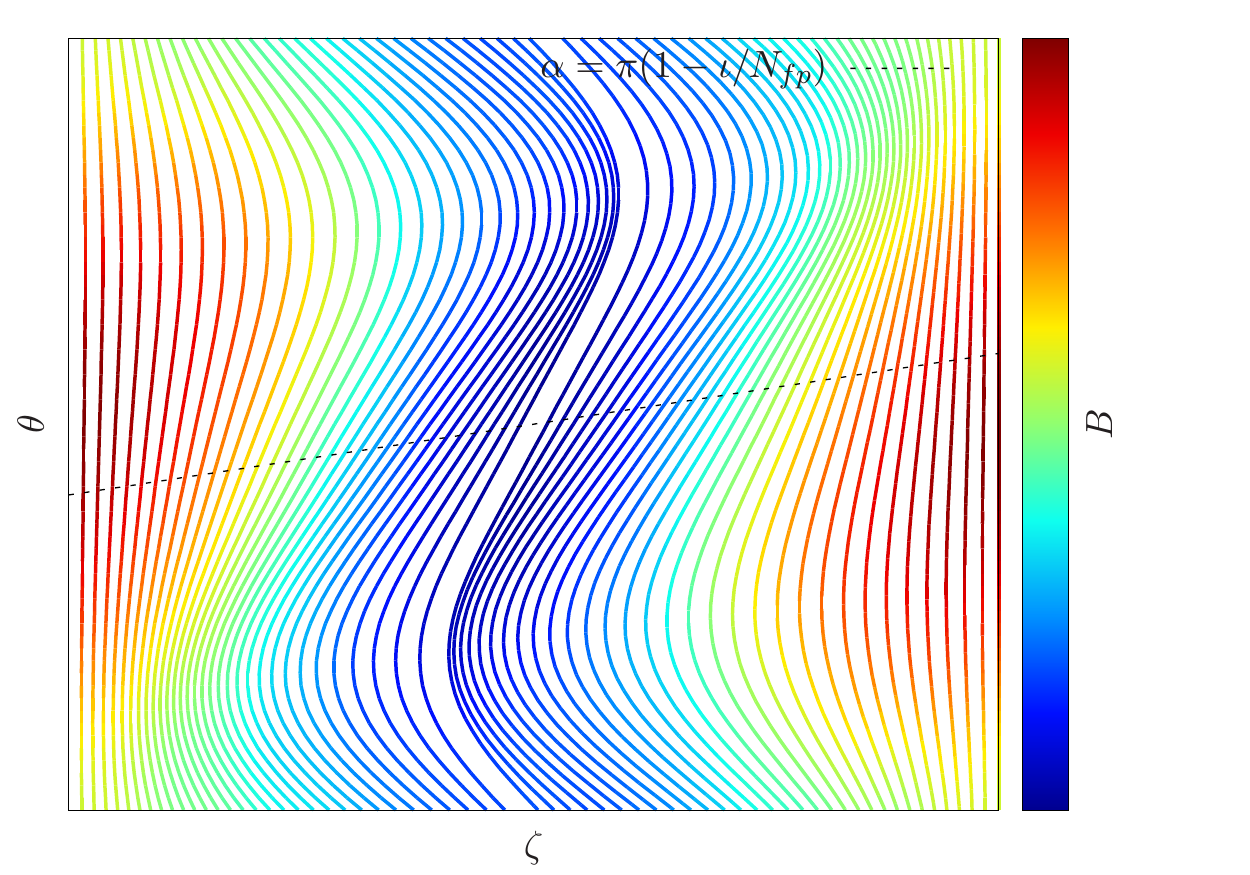}
\includegraphics[angle=0,width=0.47\columnwidth]{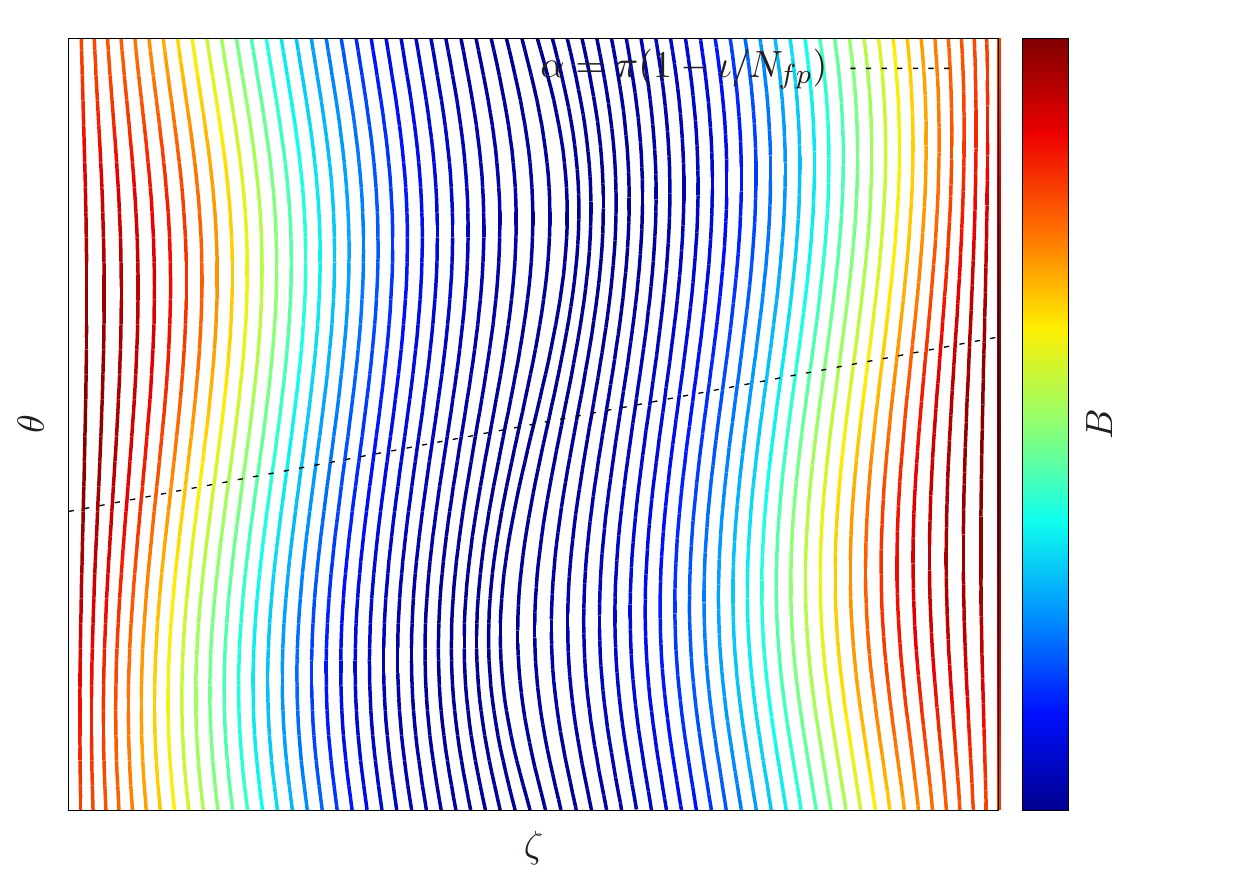}\\
\includegraphics[angle=0,width=0.47\columnwidth]{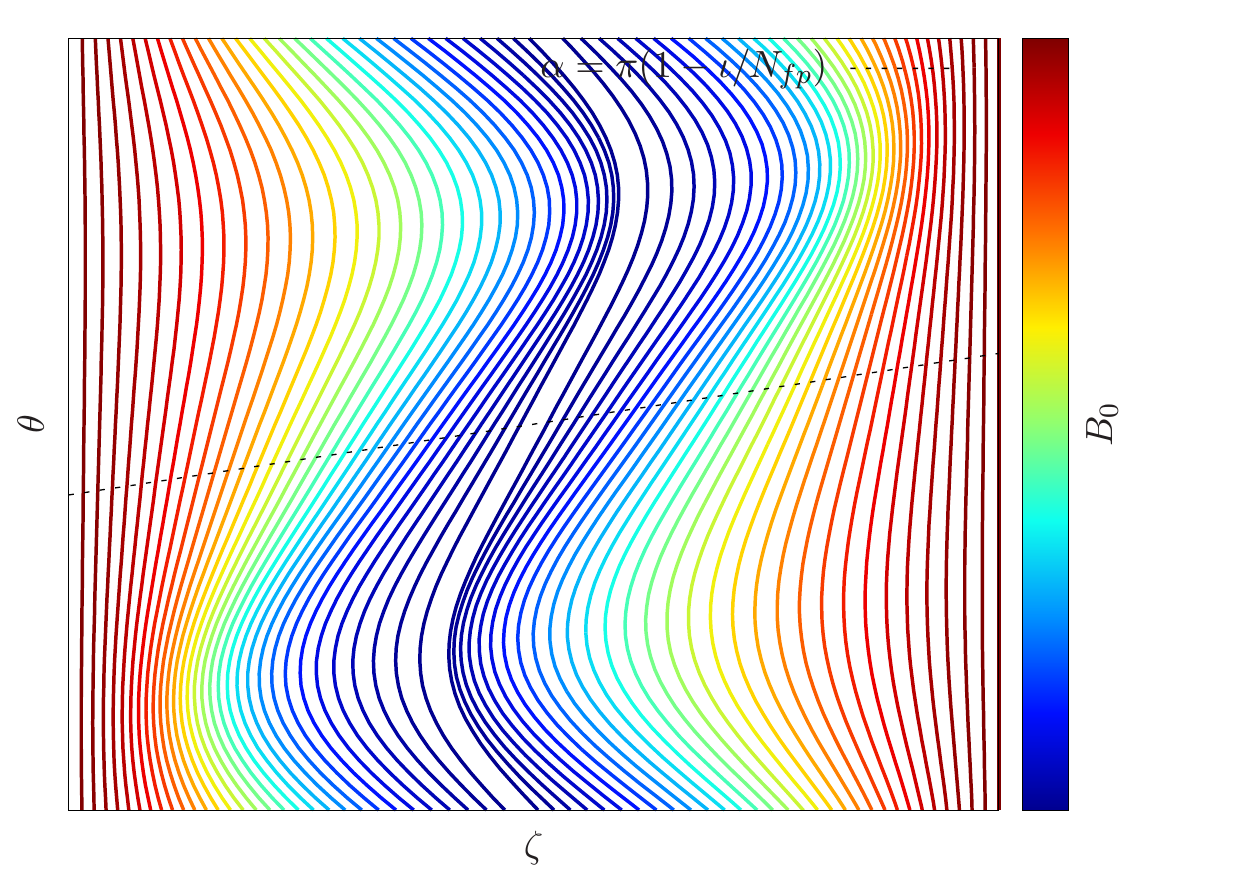}
\includegraphics[angle=0,width=0.47\columnwidth]{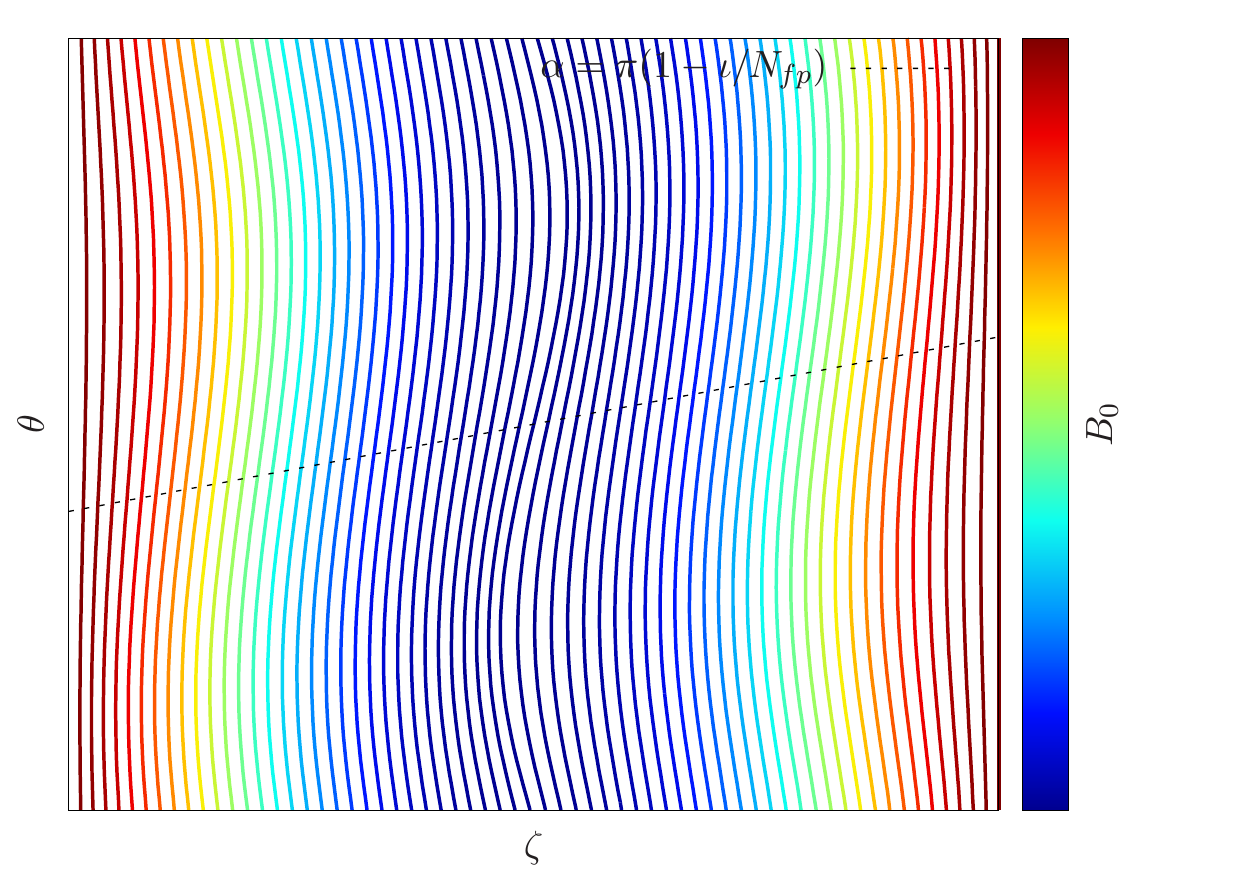}\\
\includegraphics[angle=0,width=0.47\columnwidth]{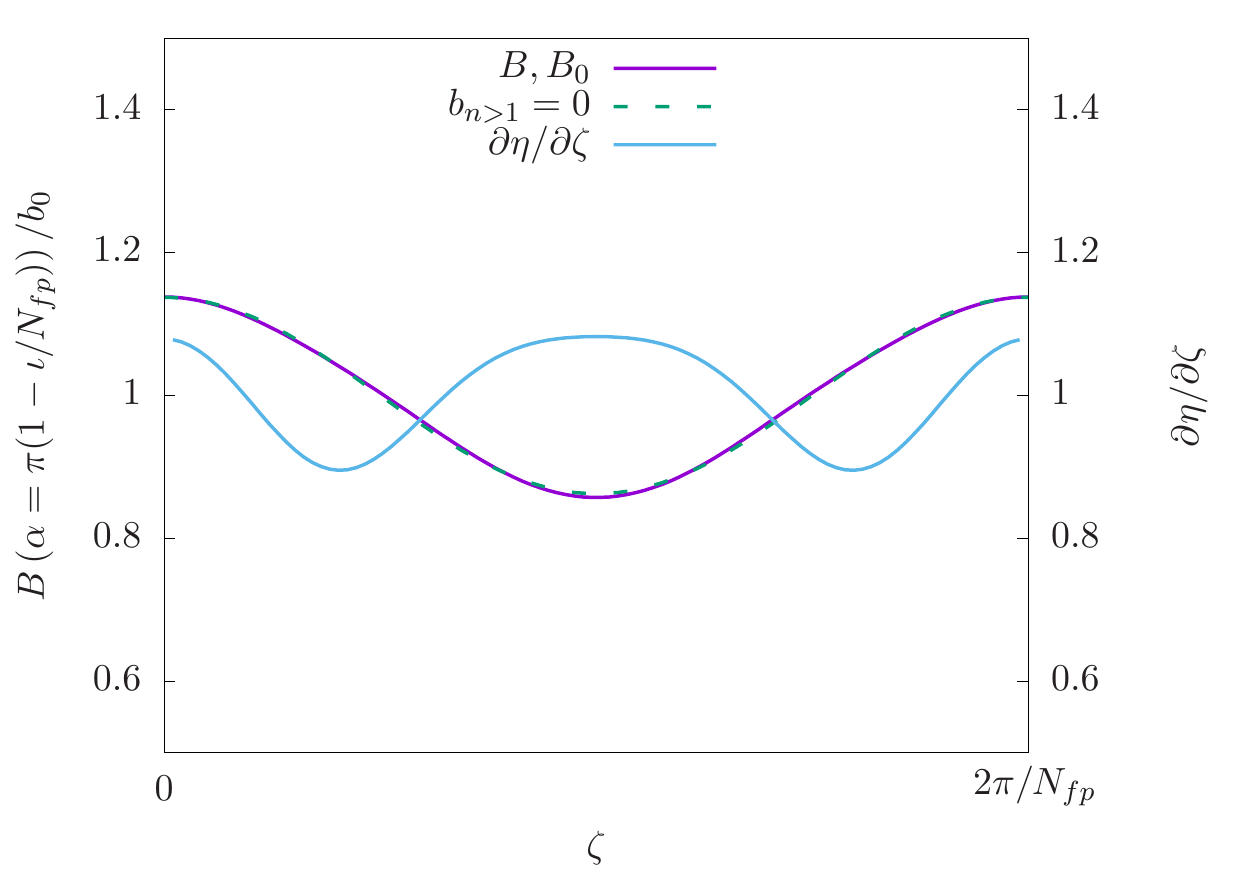}
\includegraphics[angle=0,width=0.47\columnwidth]{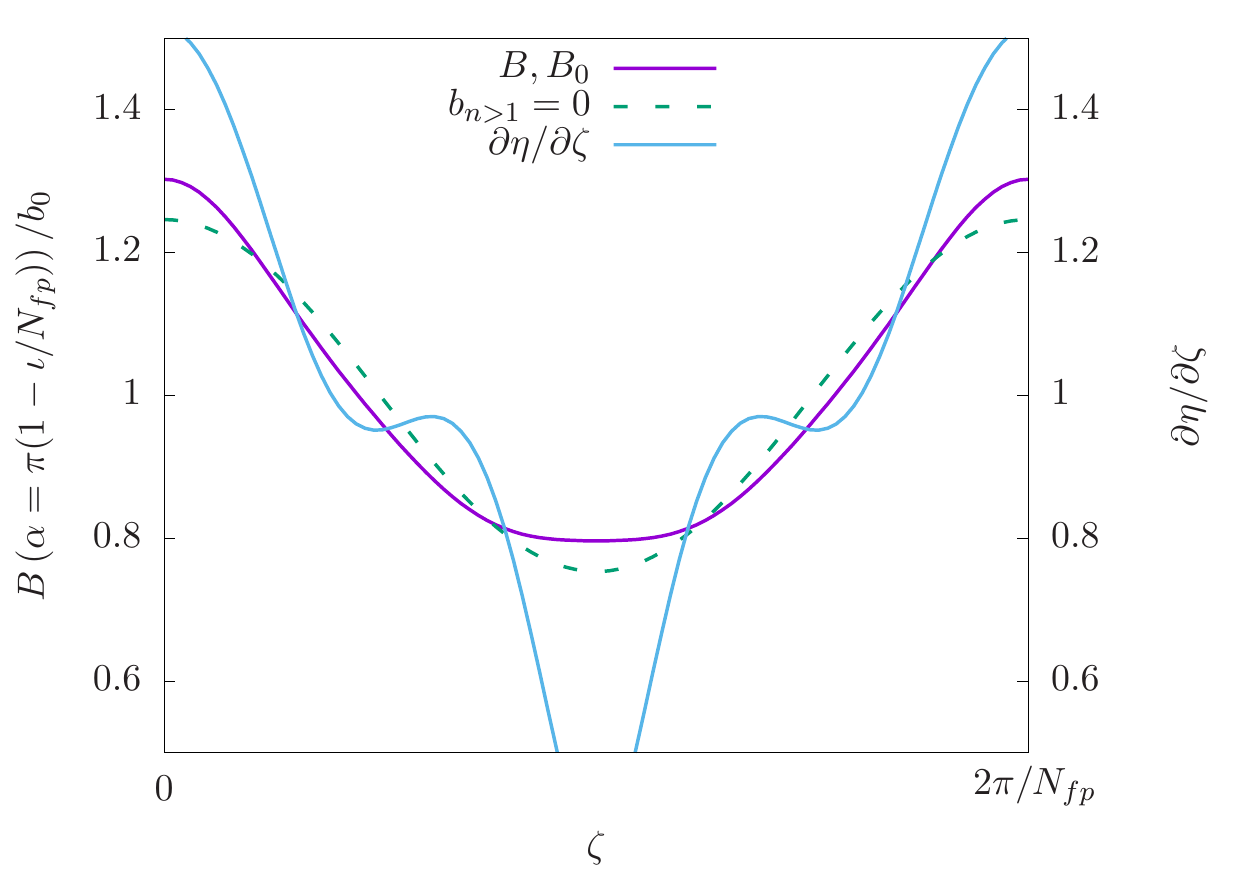}
\end{center}
\caption{For one field-period of the high mirror configuration of W7-X (left) and CIEMAT-QI (right), magnetic field strength at $s=0.3$ (top); magnetic field strength of a QI field close to the original field (center); magnetic field strength along the selected field line and value of $\partial\eta/\partial\zeta$ (bottom).}
\label{FIG_B0}
\end{figure}

\section{QI fields for which the calculation of appendix \ref{SEC_CALC} is valid}\label{SEC_B0}

In appendix~\ref{SEC_CALC}, we have employed equation (\ref{EQ_VALIDITY}) in order to simplify the calculation. In this appendix we will assess to what extent this is a valid approximation. Thanks to omnigeneity, we can focus on a particular fied-line and enforce equation (\ref{EQ_VALIDITY}) only there. By definition, $\partial_s J_0$ will take the same values for all other values of $\alpha$ of the flux-surface, even if those field-lines do not comply with equation (\ref{EQ_VALIDITY}).

Along the field line that goes through the point on the surface labelled by Boozer angles $\zeta=\pi/N_{fp}$ and $\theta=\pi$ (i.e., for  the field line for which $\alpha=\pi(1-\iota/N_{fp}))$, the magnitude of any stellarator-symmetric magnetic field has the form
\begin{equation}
B_0\left(s,\theta=\pi(1-\iota/N_{fp})+\iota\zeta,\zeta\right) = \sum_{n\ge 0} {b_n} (s) \cos(N_{fp}n\zeta)\,,\label{EQ_B0LINE} 
\end{equation}
with $b_n \neq B_{0n}$, in general. For this particular field line,
\begin{equation}
B_{00}+B_M\cos(N_{fp}\eta)=\sum_{n\ge 0} {b_n}\cos(N_{fp}n\zeta)\,,
\end{equation}  
and thus
\begin{equation}
  B_M\sin(N_{fp}\eta)\mathrm{d}\eta=\sum_{n>0}n{b_n}\sin(N_{fp} n\zeta)\mathrm{d}\zeta\,.
\end{equation}  
Using that
\begin{eqnarray}
B_{max}=B_{00}+B_M&=&\sum_{n\ge0} b_n\,,\nonumber\\
B_{min}=B_{00}-B_M&=&\sum_{n\ge0} (-1)^nb_n\,,
\end{eqnarray}
it can be obtained
\begin{equation}
\frac{\partial\eta}{\partial\zeta}=\frac{\sum_{n>0} n b_n \sin(nN_{fp}\zeta)}{\sum_{n\ge 0} b_{2n+1} \sin(N_{fp}\eta)}\,.
\end{equation}
If the field is such that
\begin{equation}
\left|\frac{n b_n}{b_1}\right| \ll 1\,,\label{EQ_COND1}
\end{equation}
for $n>1$, then
\begin{equation}
\frac{\partial\eta}{\partial\zeta}= 1+\frac{\sum_{n>1} n b_n \sin(nN_{fp}\zeta)}{b_1 \sin(N_{fp}\eta)}\,,
\end{equation}
and equation~(\ref{eq:d_sJ_3}) is correct. 

Equation (\ref{EQ_COND1}) can be fulfilled by many equilibria close to being QI. A clear example is the high-mirror configuration of W7-X, as shown in figure~\ref{FIG_B0} (left). On the other hand, this is not the case for very deeply trapped particles of fields with a broad mirror, such as CIEMAT-QI (right). We treat this case in appendix~\ref{SEC_BOTTOM}.

\section{$\partial_s J_0$ for deeply trapped particles in narrow/broad mirror QI fields}\label{SEC_BOTTOM}

Equation (\ref{EQ_COND1}), and thus equation (\ref{EQ_VALIDITY}), are not fulfilled in QI fields with a broad or a narrow mirror, since these fields cannot be described using equation (\ref{EQ_B0LINE}) with $b_n=0$ for $n>1$. In this appendix we show that, for deeply trapped trajectories, this does not change our predictions qualitatively.

Close to the minimum of $B_0$, one can Taylor-expand equation (\ref{EQ_B0LINE}) and obtain
\begin{equation}\label{EQ_BOTTOM}
B_0\left(s,\theta=\pi(1-\iota/N_{fp})+\iota\zeta,\zeta\right) \approx B_{00}-|B_{M}| + \frac{1}{2}b_w(\zeta-\pi/N_{fp})^2+ \mathcal{O}((\zeta-\pi/N_{fp})^3)\,,
\end{equation}
where $b_w(s)>0$ has been defined below equation (\ref{EQ_DSJ0BOTTOMGEN}). It is then straightforward to obtain equation (\ref{EQ_DSJ0BOTTOMGEN}). First, we compute
\begin{equation}\label{EQ_JBOTTOM}
J_0=\pi v\frac{I_p + \iota I_t}{B_{00}}\sqrt{\frac{2}{\lambda b_w}}[1-\lambda(B_{00}-|B_M|)]\,,
\end{equation}
valid in the large aspect ratio limit, and relatively large for a broad mirror, i.e., small $b_w$. Taking radial derivative and evaluating it at the minimum of $B_0$, where $\zeta=\pi/N_{fp}$ and $\lambda (B_{00}-|B_M|)=1$, yields equation (\ref{EQ_DSJ0BOTTOMGEN}).

 Restricted to the particular field line for which it was derived, equation (\ref{EQ_BOTTOM}) is valid for any stellarator-symmetric field of large aspect ratio and a single $B$-valley (exact quasi-isodynamicity is only invoked to extend its validity to other field lines of the same flux-surface). For this reason, equation (\ref{EQ_BOTTOM}) can also be employed to illustrate that exact quasi-isodynamicity is harder to obtain close to $1/\lambda=\mathcal{E}/\mu= B_{min}$, and that therefore large $|\partial_s J_0|$ is of special relevance there. Let us assume that a perturbed QI field, $B=B_0+\delta B_1$ (with $\delta\ll 1$ and $B_0\sim B_1$), can be written, for the field line given by $\alpha=\pi(1-\iota/N_{fp}))$, as
\begin{eqnarray}\label{EQ_BOTTOMP}
B\left(s,\theta=\pi(1-\iota/N_{fp})+\iota\zeta,\zeta\right)&=& B_{00}^{(0)} + \delta B_{00}^{(1)}-|B_{M}^{(0)}|-\delta |B_M^{(1)}| \nonumber\\ &+& \frac{1}{2}(b_w^{(0)}+\delta b_w^{(1)})(\zeta-\pi/N_{fp})^2\,.
\end{eqnarray}
Here, $B_{00}^{(1)}\sim B_{00}^{(0)}$, $B_{M}^{(1)}\sim B_{M}^{(0)}$, and $b_w^{(1)}\sim b_w^{(0)}$. For this field line, one obtains
\begin{equation}\
J=\pi v\frac{I_p + \iota I_t}{B_{00}^{(0)} + \delta B_{00}^{(1)}}\sqrt{\frac{2}{\lambda (b_w^{(0)}+\delta b_w^{(p)})}}[1-\lambda(B_{00}^{(0)} + \delta B_{00}^{(1)}-|B_{M}^{(0)}|-\delta |B_M^{(1)}|)]\,.
\end{equation}
To first order in $\delta$, $J$ can be written
\begin{equation}
J=J_0+\delta J_1\,,
\end{equation}
with
\begin{equation}\label{EQ_DDJ}
  \frac{J_1}{J_0}= -\frac{B_{00}^{(1)}}{B_{00}^{(0)}}-\frac{b_w^{(1)}}{2b_w^{(0)}}-\frac{\lambda(B_{00}^{(1)}-|B_M^{(1)}|)}{1-\lambda(B_{00}^{(0)}-|B_M^{(0)}|)}.
\end{equation}
This quantity can be employed to provide an estimate of the distance to perfect quasi-isodynamicity caused by a perturbation around the minimum of $B_0$. The third term of the right-hand side of equation (\ref{EQ_DDJ}) diverges at the bottom of $B_0$, which is consistent with the behaviour observed close to $1/\lambda=\mathcal{E}/\mu=B_{min}$ in figure (\ref{FIG_LOSSLAMBDA}). This can be interpreted as follows: when the bounce points along the field line in $B_0$ are very close, the relative change in $J$ caused by a given perturbation $\delta B_1$ may become very large. A very broad mirror field ($b_w^{(0)}\approx 0$ in equation (\ref{EQ_DDJ})) could contribute to reduce the size of $J_1/J_0$ (if $b_w^{(1)}$ and $B_{00}^{(1)}-|B_{M}^{(1)}|$ have opposite signs). Together with its effect on $\partial_s J_0$, a broad mirror would e.g. improve fast ion confinement, as observed in \cite{drevlak2014fastions,drevlak2018rose,sanchez2023qi}.


\bibliography{flat_mirror.bbl}

\end{document}